\documentclass[twocolumn,letterpaper, 10 pt,conference]{ieeeconf}

\IEEEoverridecommandlockouts

\usepackage{amssymb}
\usepackage{amsmath}
\usepackage{color}
\usepackage{graphics}
\usepackage{graphicx}
\usepackage{multirow}
\usepackage{cite}
\usepackage{algorithm}%,algorithmic
\usepackage{savesym}
\savesymbol{AND}
\usepackage{epstopdf}
\usepackage{subcaption}
\usepackage{float}
\usepackage{algpseudocode}

\usepackage{tikz}
\usetikzlibrary{shapes.misc}
\usepackage{circuitikz}

\def \R{\mathbb{R}}

\newtheorem{theorem}{Theorem}

\newtheorem{lemma}{Lemma}

\newtheorem{definition}{Definition}
\newtheorem{proposition}{Proposition}

\newtheorem{example}{Example}

\title{\LARGE \bf Data-driven computation of invariant sets of discrete time-invariant black-box systems}

\author{Zheming Wang and Rapha\"el M. Jungers
\thanks{The authors are with the ICTEAM Institute, UCLouvain, Louvain-la-Neuve,1348, Belgium.}
\thanks{Rapha\"el M. Jungers is a FNRS Research Associate. He is supported by the French Community of Belgium, the Walloon Region and the Innoviris Foundation.}
\thanks{Email addresses:  zheming.wang@uclouvain.be (Zheming Wang), raphael.jungers@uclouvain.be (Rapha\"el M. Jungers)}}

\begin{document}
\maketitle

\begin{abstract}
We consider the problem of computing the maximal invariant set of discrete-time black-box nonlinear systems without analytic dynamical models. Under the assumption that the system is asymptotically stable, the maximal invariant set coincides with the domain of attraction. A data-driven framework relying on the observation of trajectories is proposed to compute almost-invariant sets, which are invariant almost everywhere except a small subset. Based on these observations, scenario optimization problems are formulated and solved.
We show that probabilistic invariance guarantees on the almost-invariant sets can be established. To get explicit expressions of such sets, a set identification procedure is designed with a verification step that provides inner and outer approximations in a probabilistic sense. The proposed data-driven framework is illustrated by several numerical examples.
\end{abstract}

%%%%%%%%%%%%%%%%%%%%%%%%%%%%%%%%%%%%%%%%%%%%%%%%%%%%%%%%%%
\section{Introduction}
An invariant set of a dynamical system refers to a region where the
trajectory will never leave once it enters. It is widely used in systems and control for stability analysis and control design, see, for instance, \cite{BOO:A91,ART:Bla99,BOO:BM08} and the references therein. In particular, it has the ability to handle safety specifications and plays an important role in safety-critical applications \cite{ART:AGST13}.

Numerous algorithms have been proposed to characterize and compute invariant sets of different types of systems. The early literature has been devoted to linear systems with polyhedral constraints, see, e.g., \cite{ART:B88,ART:B88b,ART:GT91,ART:DH99}. In the presence of bounded disturbances in linear systems, robust invariant sets were studied and many algorithms have been proposed (see, e.g., \cite{ART:KG98,ART:RKKM05,ART:OG06,ART:P16}) to tackle the complications arising from the robustness requirement. Recently, the authors in  \cite{INP:WJO19}  have proposed an algorithm to deal with non-convex constraints.  Algorithms for computing invariant sets of different types of nonlinear systems can be found in \cite{INP:KM00,ART:BLAC05,ART:ACFC09,ART:FAC10,ART:KK11,ART:SG12,ART:HK14,ART:KHJ14,INC:KD15}. The concept of set invariance can also be extended to hybrid systems. For instance, the works \cite{ART:DO12,ART:HTY16,INP:AJ16,ART:ASJ17,ART:AR18} have investigated the computation of invariant sets of switching linear systems.

The algorithms in the aforementioned papers are all based on an analytic model of the system, which is usually obtained by system identification in many engineering applications \cite{INC:L98}. Most of the classical system identification methods are limited to linear systems or simple nonlinear systems. For more complex systems, piecewise affine models are often used, see, e.g., \cite{ART:BGPV05,ART:PJFV07,INP:SB18}. However, system identification for general nonlinear systems is still challenging and can often introduce considerable modeling errors. For instance, the identification problem of a switching system is known to be NP-hard \cite{ART:L16}. In view of the difficulty of system identification in real-life applications, increasing attention has been paid to data-driven analysis and control under the framework of black-box systems \cite{INP:KQHT16,INP:JSM17,ART:BFGP17,ART:KBJT19}. For instance, stability-like guarantees are provided in \cite{ART:KBJT19} for black-box switching linear systems, based merely on a finite number of observations of trajectories. Even when an analytic model is available, data-driven methods are often used for systems with complicated dynamical behavior \cite{ART:TPS08,INP:KDSA14,INP:BL15,ART:BL18} to construct Lyapunov functions and estimate the domain of attraction. As data-driven analysis only requires information obtained via simulations, it has the flexibility to deal with a broad class of complex systems. In particular, data-driven analysis allows us to study set invariance for black-box systems and design model-free algorithms for computing invariant sets of such systems. Recently, a data-driven method is presented in \cite{INP:CPGO18} to compute an approximation of a minimal robust control invariant set of linear systems with multiplicative and additive uncertainties. For nonlinear systems, the authors in \cite{INP:CRDD18} have developed an active learning method to estimate reachable and invariant sets. However, formal invariance guarantees are not provided in \cite{INP:CPGO18,INP:CRDD18}.

With only limited information via a finite number of simulations, it is very difficult to achieve an exact invariant set for general nonlinear systems. For this reason, we will use the concept of almost-invariant sets \cite{ART:DJ99} for data-driven invariance analysis. This is also closely related to the concept of almost-Lyapunov functions \cite{INP:LLZ16}, which decrease in a region almost everywhere except a small subset. Although these concepts are introduced for systems with analytic models, they are well suited for black-box systems from their definitions.

In this paper, we aim to develop a data-driven framework for computing almost-invariant sets of discrete-time black-box nonlinear system in the spirit of the scenario optimization approach \cite{ART:CC05,ART:CC06,ART:C10}, which allows us to establish probabilistic invariance guarantees. Our contribution is twofold. First, we present an algorithm that evaluates the time horizon, with which set invariance can be achieved, by using a relatively small set of trajectories with a long horizon. This will enable us to obtain an implicit representation of almost-invariant sets, based on the underlying unknown dynamics. Second, we transform this representation into an explicit one thanks to a set identification procedure, which uses a larger set of trajectories with a short horizon. We show that this step also can come with formal probabilstic guarantees and computable bounds, that can be made arbitrarily tight. Although other set identification methods can be found in \cite{INP:BNGSMW06,ART:BM10,INP:GCHK13,ART:CF18} and some can be more efficient in practice, formal guarantees are not provided in these works. Our work is the first one to provide rigorous probabilistic invariance guarantees on blackbox nonlinear systems.

%Given the data-driven nature, the proposed domain identification procedure is also suitable for the identification of general sets that are defined by the underlying dynamics in black-box systems.

%We will extend the concept of almost-invariant sets to black-box systems by using  To identify these almost-invariant sets explicitly, we also propose a data-driven procedure, which is able to generate numerical inner and outer approximation within the given confidence level. 

%\cite{ART:AGST13}

The rest of the paper is organized as follows. This section ends with the notation,  followed by the next section on the review of preliminary results on invariant sets. Section \ref{sec:timeinv} starts with the formal definition of almost-invariant sets. It then presents an intuitive approach for computing such sets. After that, we present the proposed data-driven approach. The complexity of these approaches will also be discussed in this section. In Section \ref{sec:domain}, we propose a set identification procedure to get explicit inner and outer approximations of the almost-invariant set obtained from Section \ref{sec:timeinv}. Several numerical examples are provided. The last section concludes the work.

The notation used in this paper is as follows. The non-negative integer set is indicated by $\mathbb{Z}^+$ and $\mathcal{I}_M$ denotes the set $\{1,2,\cdots, M\}$ for any $M\in \mathbb{Z}^+$.  Given sets $S$ and $Z$, the set difference $S\setminus Z$ is defined as $\{x:x \in S,x \not\in Z \}$ and $|S|$ denotes the cardinality of $S$. For two set $X$ and $Y$, $X+ Y$ denotes their Minkowski sum. Let $\lceil \cdot \rceil$ and $\lfloor \cdot \rfloor$ denote the ceil and floor functions respectively. Additional notation will be introduced as required in the text.

\section{Preliminaries}\label{sec:pre}
We consider discrete-time dynamical systems of the form
\begin{align}
x(t+1) &= f(x(t)), \quad t\in \mathbb{Z}^+, \label{eqn:fx}
\end{align}
where $x(t)\in \mathbb{R}^n$ is the state vector. The system is subject to safety constraints:
\begin{align}
x(t) &\in X, \quad t\in \mathbb{Z}^+.
\end{align}
Let us denote by $\phi(t,x)$ the solution of the system (\ref{eqn:fx}) with the initial condition $x$ at time $t=0$. In this paper, we consider the case where we do not have access to the model, i.e., to $f$ and we use the term \emph{black-box} to refer to such systems. We assume that $X\subseteq \mathbb{R}^n$ is compact and $f: \mathbb{R}^n \rightarrow \mathbb{R}^n$ is a Borel measurable function. The definition of invariant sets for the system (\ref{eqn:fx}) is given below.
\begin{definition}\label{def:cainv}
\cite{ART:Bla99} A nonempty set $Z\subseteq X$ is a positively invariant set for the system (\ref{eqn:fx}) if $f(x)\in Z$ for any $x\in Z$.
\end{definition}

Throughout this paper, all invariant sets are positively invariant sets. From the definition above, an invariant set can be considered as a safe region, where the safety constraints are always satisfied once the system enters. There often exist multiple invariant sets. In many applications, it is desirable to compute the maximal invariant set, which is defined below, as it gives us the largest safe region.

\begin{definition}\label{def:cainvmax}
\cite{ART:GT91} A nonempty set $S\subseteq X$ is the maximal invariant set for the system (\ref{eqn:fx}) if and only if $S$ is an invariant set and contains all the invariant sets in $X$.
\end{definition}

The maximal invariant set can be constructed recursively by the following iteration:
\begin{align}
O_0 := X, O_{k+1} := O_k \bigcap \{x: f(x)\in O_k\}, k \in \mathbb{Z}^+ . \label{eqn:Ok}
\end{align}
With these iterates, it can be verified that 
\begin{align}\label{eqn:Otell}
O_{k} = \{x\in X: \phi(\ell,x)\in X, \forall \ell \in \mathcal{I}_k \}
\end{align}
Thus, the maximal invariant set can be expressed as
\begin{align}\label{eqn:Oinf}
O_{\infty} = \{x\in \mathbb{R}^n: \phi(k,x)\in X, \forall k \in \mathbb{Z}^+\}.
\end{align}
From (\ref{eqn:Otell})-(\ref{eqn:Oinf}), one can see the following property:
\begin{align}\label{eqn:Onet}
X = O_0 \supseteq O_1 \supseteq \cdots \supseteq O_k \supseteq O_{k+1} \supseteq \cdots  \supseteq O_{\infty}
\end{align}
When $O_k=O_{k+1}$ for some $k\in \mathbb{Z}^+$, $O_k$ becomes the maximal invariant set, i.e., $O_k=O_{\infty}$. We will refer to the minimal $k$ satisfying the set invariance condition $O_k=O_{k+1}$ as the \emph{invariance horizon}. Under additional assumptions, the \emph{invariance horizon} is finite, see, e.g., Theorem 4.1 in \cite{ART:GT91} for linear systems. The finiteness property for nonlinear systems is stated in the following proposition.

\begin{proposition}\label{prop:Oinf}
Assume that the system (\ref{eqn:fx}) is asymptotically stable at the origin in $X$, which is compact and contains the origin in its interior, and that the function $f(x)$ is continuous with $f(0)=0$. Let $O_k$ be defined in (\ref{eqn:Ok}) for any $k\in \mathbb{Z}^+$ and the set $O_{\infty}$ be defined in (\ref{eqn:Oinf}). It has the following properties. (i) $O_{\infty}$ exists and is nonempty. (ii) If $Z\subseteq X$ is an invariant set of system (\ref{eqn:fx}), $Z\subseteq O_{\infty}$.(iii) There exists a finite $k^*$ such that
$
O_{k}=O_{k^*}
$
for all $k \ge k^*$ and $O_{\infty} = O_{k^*}$. (iv) For any $k\in \mathbb{Z}^+$, $O_k$ is compact and contains the origin in its interior. 
\end{proposition}
\textbf{Proof of Proposition \ref{prop:Oinf}}: (i) This property holds trivially since $0\in O_{\infty}$.(ii) Follows from similar arguments in \cite{ART:GT91, ART:KK11}, we can see that for any $x\in Z$, $x\in O_{\infty}$.(iii) From assumptions on asymptotic stability and compactness, there exists a $k^*$ such that $\phi(k^*+1,x)\in X$ for any $x\in X$. We claim that $O_{k^*}$ is an invariant set of system (\ref{eqn:fx}). We have to show that for any $x'\in O_{k^*}$, $f(x')\in O_{k^*}$. From the definition of $O_{k^*}$, we can see that $x\in O_{k^*}$ implies $\phi(k,x)\in X$ for all $k=0,1,\cdots, k^*$. As the system is time-invariant, we know that $\phi(k,f(x'))\in X$ for $k=0,1,\cdots, k^*-1$. From the fact that $\phi(k^*+1,x)\in X$ for any $x\in X$, we can see that $\phi(k^*,f(x'))\in X$, which implies that $f(x')\in O_{k^*}$. This means that $O_{k^*}$ is an invariant set and $O_{k^*}=O_{\infty}$. (iv) From assumptions on compactness and continuity, it can be shown that $O_k$ is closed and bounded for any finite $k\in \mathbb{Z}^+$. According to the Heine–Borel theorem, they are also compact. From the continuity of the function $f(x)$, there always exists a open ball $\mathcal{B}\in O_k$ with $0\in \mathcal{B}$ for any any finite $k\in \mathbb{Z}^+$. This completes the proof.
$\Box$

In the case where an analytic model $f$ is available, many algorithms have been proposed in \cite{ART:BLAC05,ART:FAC10,ART:KK11,ART:SG12,ART:HK14} for computing invariant sets or their approximations. However, in black-box systems, these computations are no longer feasible, since the function $f$ is unknown. In this paper, we provide a data-driven verification approach for set invariance by observing trajectories of the system with probabilistic guarantees.

\section{Main results}\label{sec:timeinv}
%\begin{definition}
%For some $\epsilon>0$, the nonempty set $Z\subseteq X$ is a $\epsilon$-\emph{CA-invariant} set for system (\ref{eqn:ftx}) if and only if $Pr(\phi(k,x)\not\in Z, \textrm{ for some } k\in \mathbb{Z}^+|x\in Z)\le \epsilon$.
%\end{definition}

%\subsection{Linear time-varying systems}
%A linear time-varying system is a dynamical system of the form (\ref{eqn:ftx}), with $f(t,x(t))=A(t)x(t)$, that is:
%\begin{align}
%x(t+1) = A(t)x(t), \quad t \in \mathbb{Z}^+
%\end{align}
%where $A(t): \mathbb{Z}^+ \rightarrow \mathbb{R}^{n\times n}$ is a time-varying matrix. An additional assumption is made.
%
%\begin{assumption}\label{ass:polytope}
%The set $X$ is a polytope and contains the origin in its interior, with $p$ vertices $V:=\{v_1,v_2,\cdots, v_p\}$.
%\end{assumption}
%
%With this assumption, we only need to trace the trajectories starting from the vertices and a modification of problem (\ref{eqn:tmin}) is given below
%
%\begin{subequations}\label{eqn:tminV}
%\begin{align}
%&\min_{t\in \mathbb{Z}^+} t\\
%\mathrm{s.t.} \quad &\phi(k+1,v) \in X, \forall v\in V, \forall k\ge t.
%\end{align}
%\end{subequations}
%
%This problem is computationally tractable since we can let the system propagate for a sufficient long time window starting from these vertices.

In this section, we will discuss set invariance verification under the data-driven framework. We will focus on almost-invariant sets, where the size of the invariance violating subset can be made arbitrarily small.

%To obtain such sets, we first present an intuitive approach, which estimates the size of the set defined in (\ref{eqn:Otell}) in a Monte Carlo fashion, with probabilistic set invariance guarantees. Then, we modify this approach by estimating the invariance horizon directly and derive an improved bound.

\subsection{Almost-invariant sets}
Without any dynamical model, we attempt to verify set invariance by observing $N$ trajectories that are generated from $N$ initial conditions in $X$.  Consider the Borel $\sigma$-algebra on $X$, denoted by $\mathcal{F}$, and the uniform probability measure $\mathbb{P}: \mathcal{F} \rightarrow [0,1]$, these initial conditions, denoted by $\omega_N=\{x_1,x_2,\cdots, x_N\}$, are sampled randomly from the set $X$ according to $\mathbb{P}$. More formally, the sample $\omega_N$ is independent and identically distributed (i.i.d.) with respect to the uniform distribution $\mathbb{P}$ on $X$.  For each initial condition $x\in \omega_N$, we will generate a trajectory by letting the system propagate for a sufficiently long time.

We consider sets that are almost invariant except in an arbitrarily small subset. Such a set is referred to as an almost-invariant set, which is formally defined below, adapted from \cite{ART:DJ99}.

\begin{definition}\label{def:almost}
For any $\epsilon\in [0,1]$, the set $Z\in \mathcal{F}$ is an $\epsilon$ almost-invariant set for the system (\ref{eqn:fx}) if $\mathbb{P}( \{x\in Z: f(x)\not\in Z\}) \le \epsilon$.
\end{definition}

As $f(x)$ is Borel measurable, it can be shown that $\{x\in Z: f(x)\not\in Z\} = Z\setminus \{x\in Z: f(x)\in Z\} \in \mathcal{F}$. Thus, almost-invariant sets are well defined. For the verification of the set invariance property, let us define:
\begin{align}\label{eqn:St}
S(k) := \mathbb{P}(O_{k} \setminus O_{k+1}), \forall k\in \mathbb{Z}^+.
\end{align}
From (\ref{eqn:Ok}), it is not difficult to see that $S(k)=\mathbb{P}( \{x\in O_{k}: f(x)\not\in O_{k}\})$ for all $k\in \mathbb{Z}^+$. Hence, given an $\epsilon$, $O_t$ is an $\epsilon$ almost-invariant set for the system (\ref{eqn:fx}) when $S(k)\le \epsilon$ for some $k\in \mathbb{Z}^+$.

\subsection{Data-driven set invariance verification}\label{sec:intuitive}
With the definition of almost-invariant sets, we now discuss data-driven verification for the set invariance condition $O_k=O_{k+1}$ in a Monte Carlo fashion. More precisely, we count the number of points inside $O_{k}$ given a total number of sampled initial conditions and the ratio will be an estimate of the measure $\mathbb{P}(O_{k})$ for all $k\in \mathbb{Z}^+$. Let us define the indicator function of $O_k$, $\forall k\in \mathbb{Z}^+$,
\begin{align}
\pmb{1}_{O_{k}}(x) = \begin{cases}
      1 & \textrm{If } x\in X, \textrm{ and } \phi(\ell,x)\in X, ~ \forall \ell \in \mathcal{I}_k\\
      0 & \textrm{otherwise}.\\
   \end{cases}
\end{align}
Consider $N$ random initial states $\omega_N$, let
\begin{align}\label{eqn:thetaN}
\theta_k(\omega_N) = \frac{\sum\limits_{x\in \omega_N}\pmb{1}_{O_{k}}(x)}{N}, \quad k \in \mathbb{Z}^+.
\end{align}
With the sequence of the estimates above, we then define:
\begin{align}\label{eqn:ttheta}
\bar{t}(\omega_N) : = \min_{k \in \mathbb{Z}^+} k: \theta_k(\omega_N) =\theta_{k+1}(\omega_N) 
\end{align}
As $N$ is finite and $\{\theta_k(\omega_N)\}_{k\in \mathbb{Z}^+}$ is a non-increasing sequence, $\bar{t}(\omega_N)$ is always finite. The condition in (\ref{eqn:ttheta}) can be considered as a data-driven version of the white-box set invariance condition $O_{k}=O_{k+1}$. When the set $\omega_N$ is sufficiently large, $O_{\bar{t}(\omega_N)}$ can be arbitrarily close to the real maximal invariant set with arbitrarily high probability. This is formally stated in the following theorem.
\begin{theorem}\label{thm:thetaN}
Given $N$ random initial states $\omega_N$ extracted according to the distribution $\mathbb{P}$ over $X$ and any $k \in \mathbb{Z}^+$, let $\theta_k(\omega_N)$ be defined as in (\ref{eqn:thetaN}). $\mathbb{P}^N$ denotes the probability measure in the space $X^N$ of the multi-sample extraction $\omega_N$. Let $O_k$ and $S(k)$ be defined by (\ref{eqn:Ok}) and (\ref{eqn:St}). Then, for any $\epsilon\in (0,1]$, 
\begin{align}
\mathbb{P}^N(\omega_N\in X^N:S(\bar{t}(\omega_N))\ge \epsilon) &\le \frac{1}{\epsilon} (1-\epsilon)^N   \label{eqn:stepsilontheta}
\end{align}
%\begin{align}
%\mathbb{P}^N(\omega_N\in X^N:|\theta_k(\omega_N)-\gamma_k | \ge \epsilon)& \le 2e^{-2N\epsilon^2},  \label{eqn:thetagammabound}\\
%\mathbb{P}^N(\omega_N\in X^N:S(\bar{t}(\omega_N))\ge \epsilon) &\le 4e^{-\frac{1}{2}N\epsilon^2}  \label{eqn:stepsilontheta}
%\end{align}
where $\bar{t}(\omega_N)$ is defined as in (\ref{eqn:ttheta}).
\end{theorem}
\textbf{Proof of Theorem \ref{thm:thetaN}}:
Given any $\epsilon\in (0,1]$, we consider the set $\mathcal{I}_{\epsilon} = \{k\in \mathbb{Z}^+: S(k) \ge \epsilon\}$. The set $\{\omega_N\in X^N:S(\bar{t}(\omega_N))\ge \epsilon\}$ can be written as $\cup_{k\in \mathcal{I}_{\epsilon} }\{\omega_N\in X^N: \bar{t}(\omega_N) = k \}$. For each $k\in \mathcal{I}_{\epsilon}$, $\{\omega_N\in X^N: \bar{t}(\omega_N) = k \}$ is a subset of $\{\omega_N\in X^N: \theta_k(\omega_N) =\theta_{k+1}(\omega_N) \} = \{\omega_N\in X^N: \omega_N \cap (O_k\setminus O_{k+1})=\emptyset \}$, whose measure is bounded by $(1-\epsilon)^N$ as $S(k) \ge \epsilon$. It is easy to verify that the cardinality of $\mathcal{I}_{\epsilon}$ is bounded by $\frac{1}{\epsilon}$. Hence, the measure of the set  $\{\omega_N\in X^N:S(\bar{t}(\omega_N))\ge \epsilon\}$ is bounded by $\frac{1}{\epsilon} (1-\epsilon)^N$. $\Box$

From Theorem \ref{thm:thetaN}, one can see that, for an given $\epsilon\in (0,1]$, the probability of the event that $S(\bar{t}(\omega_N))\le \epsilon$ is at least $1-\frac{1}{\epsilon} (1-\epsilon)^N$. From Definition \ref{def:almost}, for any confidence parameter $\beta\in (0,1]$, $O_{\bar{t}(\omega_N)}$ is an $\epsilon$ almost-invariant set of the system (\ref{eqn:fx}) with probability no smaller than $1-\beta$ if $N\ge \frac{\ln(\epsilon\beta)}{\ln(1-\epsilon)}$. As a result, we need to take $\lceil \frac{\ln(\epsilon\beta)}{\ln(1-\epsilon)} \rceil$ initial states to get an a priori guaranteed $\epsilon$ almost-invariant set with probability at least $1-\beta$ for the given $\epsilon$ and $\beta$.

\subsection{An improved bound}
With modifications on the estimation of the \emph{invariance horizon} in (\ref{eqn:ttheta}), the bound in Theorem \ref{thm:thetaN} can be significantly improved for reasonable tolerances and confidence levels. For any initial state $x\in X$, let us denote the first time that the system leaves the constraint set $X$ by
\begin{align}\label{eqn:tstarxnot}
t^*(x) := \min_{t\in \mathbb{Z}^+} t: \phi(t,x)\not\in X,
\end{align}
and $t^*(x)=0$ when $\phi(t,x)\in X$ for all $t\in \mathbb{Z}^+$.  With slight abuse of notation, we define:
\begin{align}\label{eqn:Xphiomega}
 t^*(\omega_N):=\max\limits_{x\in \omega_N} t^*(x)
\end{align}
for the given the initial states $\omega_N$. The following lemma shows that (\ref{eqn:Xphiomega}) leads to a stronger data-driven condition.

\begin{lemma}\label{lem:tsaromega}
Given $N$ initial states, denoted by $\omega_N$, let $\theta_k(\omega_N)$ be defined in (\ref{eqn:thetaN}) for $k\in \mathbb{Z}^+$ and $t^*(\omega_N)$ be defined in (\ref{eqn:Xphiomega}). It holds that
\begin{align}
t^*(\omega_N) = \min_{k \in \mathbb{Z}^+} k: \theta_k(\omega_N) =\theta_{\infty}(\omega_N).
\end{align}
\end{lemma}
Proof of Lemma \ref{lem:tsaromega}: This result is direct from the definitions of $t^*(\omega_N)$ and $\theta_k(\omega_N) $. $\Box$

From the lemma above, $t^*(\omega_N) \ge \bar{t}(\omega_N)$ for the same $\omega_N$, which is illustrated in Figure \ref{fig:tbartstar}. With (\ref{eqn:Xphiomega}), an improved bound can be derived as stated in the following theorem.

\tikzset{cross/.style={cross out, draw, 
         minimum size=2*(#1-\pgflinewidth), 
         inner sep=0pt, outer sep=0pt}}
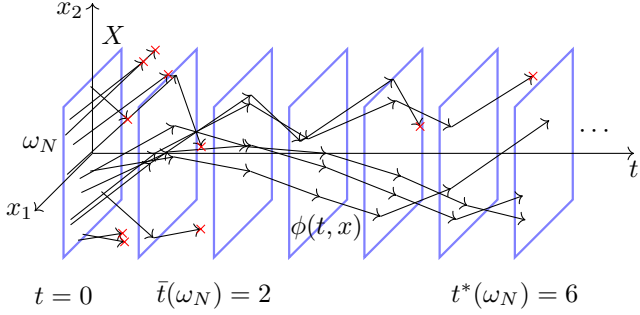
\begin{figure}[h]
\centering
\begin{tikzpicture}[draw=black]
\draw[line width=0.3mm,blue,fill=white,opacity=0.5] (0,0,0) -- ++(0,0,-2) -- ++(0,-2,0) -- ++(0,0,2) -- cycle;

\draw[line width=0.3mm,blue,fill=white,opacity=0.5] (1,0,0) -- ++(0,0,-2) -- ++(0,-2,0) -- ++(0,0,2) -- cycle;

\draw[line width=0.3mm,blue,fill=white,opacity=0.5] (2,0,0) -- ++(0,0,-2) -- ++(0,-2,0) -- ++(0,0,2) -- cycle;

\draw[line width=0.3mm,blue,fill=white,opacity=0.5] (3,0,0) -- ++(0,0,-2) -- ++(0,-2,0) -- ++(0,0,2) -- cycle;

\draw[line width=0.3mm,blue,fill=white,opacity=0.5] (4,0,0) -- ++(0,0,-2) -- ++(0,-2,0) -- ++(0,0,2) -- cycle;

\draw[line width=0.3mm,blue,fill=white,opacity=0.5] (5,0,0) -- ++(0,0,-2) -- ++(0,-2,0) -- ++(0,0,2) -- cycle;

\draw[line width=0.3mm,blue,fill=white,opacity=0.5] (6,0,0) -- ++(0,0,-2) -- ++(0,-2,0) -- ++(0,0,2) -- cycle;

\draw[->,draw=black] (0,-1,-1) -- ++(7.2,0,0);
\draw[->,draw=black] (0,-1,-1) -- ++(0,2,0);
\draw[->,draw=black] (0,-1,-1) -- ++(0,0,2);
\draw(7.2,-1.2,-1) node {$t$};
\draw(-0.2,-1,1) node {$x_1$};
\draw(-0.3,0.9,-1) node {$x_2$};
\draw(6.7,-0.7,-1) node {$\cdots$};

\draw[->] (0,-1.1660,-0.5594) -- (1,-0.7253,-1.2425);
\draw[->] (1,-0.7253,-1.2425) -- (2,-0.9678,-1.1467);
\draw[->] (2,-0.9678,-1.1467) -- (3,-1.1145,-1.2955);
\draw[->] (3,-1.1145,-1.2955) -- (4,-1.4100,-1.2714);
\draw[->] (4,-1.4100,-1.2714) -- (5,-1.6814,-0.9613);
\draw[->] (5,-1.6814,-0.9613) -- (6,-1.6427,-0.3689);

%\draw[->,draw=red] (0,-1.9998,-1.3953) -- (1,-2.3951,0.7015);

%\draw[->,draw=red] (0,-1.7065,-1.8153) -- (1,-2.5218,-0.2993);

\draw[->] (0,-1.6275,-1.3089) -- (1,-1.9364,-0.5101);
\draw[->] (1,-1.9364,-0.5101) -- (2,-1.4465,0.4559);
\draw(2,-1.4465,0.4559) node[cross=2pt,draw=red] {};

\draw[->] (0,-1.2065,-0.9224) -- (1,-1.1288,-1.2123);
\draw[->] (1,-1.1288,-1.2123) -- (2,-1.3412,-1.2644);
\draw[->] (2,-1.3412,-1.2644) -- (3,-1.6056,-1.0648);
\draw[->] (3,-1.6056,-1.0648) -- (4,-1.6704,-0.5641);
\draw[->] (4,-1.6704,-0.5641) -- (5,-1.2345,-0.3988);
\draw[->] (5,-1.2345,-0.3988) -- (6,-0.6333,-1.1876);

%\draw[->] (0,-1.1616,-0.6296) -- (1,-0.7912,-1.2454);
%\draw[->] (1,-0.7912,-1.2454) -- (2,-1.0365,-1.2104);
%\draw[->] (2,-1.0365,-1.2104) -- (3,-1.2469,-1.2949);
%\draw[->] (3,-1.2469,-1.2949) -- (4,-1.5418,-1.1757);
%\draw[->] (4,-1.5418,-1.1757) -- (5,-1.7175,-0.7104);
%\draw[->] (5,-1.7175,-0.7104) -- (6,-1.4279,-0.2680);

\draw[->] (0,-1.5911,-0.2438) -- (1,-0.8349,-0.5988);
\draw[->] (1,-0.8349,-0.5988) -- (2,-0.4337,-1.2431);
\draw[->] (2,-0.4337,-1.2431) -- (3,-0.6767,-0.6561);
\draw[->] (3,-0.6767,-0.6561) -- (4,-0.3329,-1.0886);
\draw[->] (4,-0.3329,-1.0886) -- (5,-0.4215,-0.4075);
\draw[->] (5,-0.4215,-0.4075) -- (6,0.1710,-0.6282);
\draw(6,0.1710,-0.6282) node[cross=2pt,draw=red] {};

\draw[->] (0,-1.9452,-0.6591) -- (1,-1.6043,0.4893);
\draw(1,-1.6043,0.4893) node[cross=2pt,draw=red] {};

%\draw[->] (0,-1.1654,-0.8826) -- (1,-1.0480,-1.2429);
%\draw[->] (1,-1.0480,-1.2429) -- (2,-1.2909,-1.2930);
%\draw[->] (2,-1.2909,-1.2930) -- (3,-1.5839,-1.1284);
%\draw[->] (3,-1.5839,-1.1284) -- (4,-1.7122,-0.6157);
%\draw[->] (4,-1.7122,-0.6157) -- (5,-1.3280,-0.2830);
%\draw[->] (5,-1.3280,-0.2830) -- (6,-0.6110,-1.0825);

%\draw[->,draw=red] (0,-1.7192,-1.6038) -- (1,-2.3230,-0.2630);

\draw[->] (0,-0.3985,-0.0635) -- (1,0.5380,-0.5740);
\draw(1,0.5380,-0.5740) node[cross=2pt,draw=red] {};

\draw[->] (0,-1.3732,-0.6154) -- (1,-0.9885,-1.0191);
\draw[->] (1,-0.9885,-1.0191) -- (2,-1.0076,-1.2973);
\draw[->] (2,-1.0076,-1.2973) -- (3,-1.3050,-1.2975);
\draw[->] (3,-1.3050,-1.2975) -- (4,-1.6024,-1.1116);
\draw[->] (4,-1.6024,-1.1116) -- (5,-1.7140,-0.5717);
\draw[->] (5,-1.7140,-0.5717) -- (6,-1.2858,-0.2779);

\draw[->] (0,-0.2472,-0.2108) -- (1,0.5420,-0.1642);
\draw(1,0.5420,-0.1642) node[cross=2pt,draw=red] {};

%\draw[->,draw=red] (0,-1.8299,-1.9219) -- (1,-2.7518,0.0799);

\draw[->] (0,-1.6603,-0.2437) -- (1,-0.9041,-0.4255);
\draw[->] (1,-0.9041,-0.4255) -- (2,-0.3296,-1.2792);
\draw[->] (2,-0.3296,-1.2792) -- (3,-0.6087,-0.3986);
\draw[->] (3,-0.6087,-0.3986) -- (4,-0.0074,-0.9914);
\draw[->] (4,-0.0074,-0.9914) -- (5,0.0012,0.6731);
\draw(5,0.0012,0.6731) node[cross=2pt,draw=red] {};

%\draw[->] (0,-1.8033,-1.1578) -- (1,-1.9611,-0.0070);
%\draw[->,draw=red] (1,-1.9611,-0.0070) -- (2,-0.9681,0.5498);

\draw[->] (0,-0.9442,-0.1249) -- (1,-0.0690,-1.2914);
\draw[->] (1,-0.0690,-1.2914) -- (2,-0.3604,0.4358);
\draw(2,-0.3604,0.4358) node[cross=2pt,draw=red] {};

%\draw[->] (0,-0.9566,-1.7836) -- (1,-1.7402,-1.2938);
%\draw[->,draw=red] (1,-1.7402,-1.2938) -- (2,-2.0341,-0.2017);

\draw[->] (0,-0.0842,-0.9337) -- (1,-0.0179,0.3797);
\draw(1,-0.0179,0.3797) node[cross=2pt,draw=red] {};

%\draw[->] (0,-0.6162,-1.3690) -- (1,-0.9852,-1.0031);
%\draw[->] (1,-0.9852,-1.0031) -- (2,-0.9883,-1.2972);
%\draw[->] (2,-0.9883,-1.2972) -- (3,-1.2854,-1.2973);
%\draw[->] (3,-1.2854,-1.2973) -- (4,-1.5828,-1.1346);
%\draw[->] (4,-1.5828,-1.1346) -- (5,-1.7174,-0.6184);
%\draw[->] (5,-1.7174,-0.6184) -- (6,-1.3358,-0.2682);

\draw[->] (0,-0.6270,-0.3307) -- (1,0.0423,-1.0193);
\draw(1,0.0423,-1.0193) node[cross=2pt,draw=red] {};

\draw[->] (0,-1.9634,-0.4997) -- (1,-1.4631,0.5588);
\draw(1,-1.4631,0.5588) node[cross=2pt,draw=red] {};

%\draw(0,-0.7253,-1.2425) node[cross] {};
%\draw(0,-0.9678,-1.1467) node[cross] {};
%\draw(0,-1.1145,-1.2955) node[cross] {};
%\draw(0,-1.4100,-1.2714) node[cross] {};
%\draw(0,-1.6814,-0.9613) node[cross] {};
%\draw(0,-1.6427,-0.3689) node[cross] {};

\draw(0,-2.5,0) node {$t=0$};
\draw(2,-2.5,0) node {$\bar{t}(\omega_N)=2$};
\draw(6,-2.5,0) node {$t^*(\omega_N)=6$};
\draw(-0.3,0,-2.5) node {$X$};
\draw(3.5,-1.6,0) node {$\phi(t,x)$};
\draw(-0.3,-0.5,0) node {$\omega_N$};
\end{tikzpicture}
\caption{Illustration of $\bar{t}(\omega_N)$ and $t^*(\omega_N)$: the red cross denotes $t^*(x)$, i.e., the first time that the system leaves $X$.}
\label{fig:tbartstar}
\end{figure}

\begin{theorem}\label{thm:EStN1}
For all $k\in \mathbb{Z}^+$, let $O_k$ be defined by (\ref{eqn:Ok}). Given $N\in \mathbb{Z}^+$ and $N$ random initial states $\omega_N$ extracted according to the distribution $\mathbb{P}$ over $X$, let $t^*(\omega_N)$ be given in (\ref{eqn:Xphiomega}) and $S(k)$ be defined in (\ref{eqn:St}) for all $k\in \mathbb{Z}^+$. $\mathbb{P}^N$ denotes the probability measure in the space $X^N$ of the multi-sample extraction $\omega_N$. Then, for any $\epsilon\in (0,1]$, 
\begin{align}
\mathbb{P}^N(S(t^*(\omega_N))\ge \epsilon) &\le \mathbb{P}^N(\mathbb{P}(O_{t^*(\omega_N)}\setminus O_{\infty})\ge \epsilon) \nonumber\\
&\le (1-\epsilon)^N. \label{eqn:PNepsilon1}
\end{align}
\end{theorem}
\textbf{Proof of Theorem \ref{thm:EStN1}}:
 First, we show that $\mathbb{P}^N(\mathbb{P}(O_{t^*(\omega_N)}\setminus O_{\infty})\ge \epsilon)\le (1-\epsilon)^N$. For a given $\epsilon$, we consider two cases. Case one: $\mathbb{P}(O_k\setminus O_{\infty})< \epsilon$ for all $k\in \mathbb{Z}^+$. In this case, it holds trivially that $\mathbb{P}^N(\mathbb{P}(O_{t^*(\omega_N)}\setminus O_{\infty})\ge \epsilon)=0\le (1-\epsilon)^N$. Case two: there exists a $k$ such that $\mathbb{P}(O_k\setminus O_{\infty})\ge  \epsilon$. From the convergence of $\{O_{k}\}_{k\in \mathbb{Z}^+}$, there always exists $k'$ such that $\mathbb{P}(O_{k'}\setminus O_{\infty})\ge  \epsilon$ and $\mathbb{P}(O_{k}\setminus O_{\infty})<  \epsilon$ for all $k> k'$. Hence, $\{\omega_N: t^*(\omega_N)\ge k'+1\}=\{\omega_N: \mathbb{P}(O_{t^*(\omega_N)}\setminus O_{\infty})< \epsilon\}$. Considering the fact that $t^*(x) = k+1$ if and only if $x\in O_k\setminus O_{k+1}$ for any $k\in \mathbb{Z}^+$, we know that $t^*(\omega_N)\ge k'+1$ if and only if there exists at least one sampled point inside $O_{k'}\setminus O_{\infty}$. Since $\mathbb{P}(O_{k'}\setminus O_{\infty})\ge  \epsilon$, the measure of the set $\{\omega_N: t^*(\omega_N)\ge k'+1\}$ is bounded from below by $1-(1-\epsilon)^N$. Therefore, $\mathbb{P}^N(\mathbb{P}(O_{t^*(\omega_N)}\setminus O_{\infty})\ge \epsilon) \le (1-\epsilon)^N$. The inequality $\mathbb{P}^N(S(t^*(\omega_N))\ge \epsilon) \le \mathbb{P}^N(\mathbb{P}(O_{t^*(\omega_N)}\setminus O_{\infty})\ge \epsilon)$ holds because $S(t^*(\omega_N))\le \mathbb{P}(O_{t^*(\omega_N)}\setminus O_{\infty})$, which implies $\{\omega_N: S(t^*(\omega_N)) \ge \epsilon\}\subseteq \{\omega_N: \mathbb{P}(O_{t^*(\omega_N)}\setminus O_{\infty}) \ge \epsilon\}$. From Lemma \ref{lem:tsaromega}, we can see that $\theta_{t^*(\omega_N)}(\omega_N)=\theta_{t^*(\omega_N)+1}(\omega_N)$. Then, from Theorem \ref{thm:thetaN}, we can also get $\mathbb{P}^N(S(t^*(\omega_N))\ge \epsilon) \le 4e^{-\frac{1}{2}N\epsilon^2}$, which is quite loose compared with the bound above.
$\Box$

From (\ref{eqn:PNepsilon1}), one can see that, for a given $\epsilon\in (0,1]$ and a confidence parameter $\beta\in (0,1]$, if $N \ge \frac{\ln(\beta)}{\ln(1-\epsilon)}$, $O_{t^*(\omega_N)}$ is an $\epsilon$ almost-invariant set of the system (\ref{eqn:fx}) with probability no smaller than $1-\beta$. Hence, with the improved bound in Theorem \ref{thm:EStN1}, we only need to sample $\lceil\frac{\ln(\beta)}{\ln(1-\epsilon)}\rceil$ initial states for the same $\epsilon$ and $\beta$. 

It is straightforward to compute $\bar{t}(\omega_N)$, while the computation of $t^*(\omega_N)$ is more complicated. For general nonlinear systems, given a point $x\in \mathbb{R}^n$, we cannot  verify explicitly that $t^*(x)=0$, i.e., $\phi(t,x)\in X$ for all $t\in \mathbb{Z}^+$. Hence, a stopping criterion is used: $t^*(x)$ is considered to be $0$, if the system starting from the initial state $x$ does not leave $X$ for a long horizon and the state stays close to the past trajectory. The procedure for computing $\bar{t}(\omega_N)$ and $t^*(\omega_N)$ is described in Algorithm \ref{algo:nltimeinv}.

\begin{algorithm}[h]
\caption{Invariance horizon estimation for black-box systems}
\begin{algorithmic}[1]
\renewcommand{\algorithmicrequire}{\textbf{Input:}}
\renewcommand{\algorithmicensure}{\textbf{Output:}}
\Require $X$, $\omega_N$, $\delta$ (tolerance), and $\bar{T}$ (a long horizon)
\Ensure  $\bar{t}(\omega_N)$, $t^*(\omega_N)$
\\\textit{Initialization}: Set $T\leftarrow 0$, $\Delta \leftarrow 0 $, $\bar{t}(\omega_N) \leftarrow  0$, $t^*(\omega_N) \leftarrow  0$, $\Omega \leftarrow \omega_N$ and $\bar{\Omega} \leftarrow  \emptyset$;
\For{ all $x$ inside $\Omega$}
\State  Generate a new state $\phi(T+1,x)$;
\State If the new state $\phi(T+1,x)$ leaves $X$, let $t^*(\omega_N) \leftarrow  T+1$ and $\bar{\Omega} \leftarrow  \bar{\Omega} \cup \{x\}$; otherwise, compute the distance to the previous trajectory  and let
$\Delta \leftarrow  \max\{\Delta,\min_{k\le T}\|\phi(T+1,x)-\phi(k,x)\|\}$;
\EndFor
\State Case 1: $\bar{\Omega}=\emptyset$. Let $\bar{t}(\omega_N)\leftarrow  T$ (when $\bar{t}(\omega_N)$ has not been updated yet, i.e., $\bar{t}(\omega_N)=0$). Then, check $\Delta$ (the longest distance to the previous trajectory over all the points in $\Omega$) and the current horizon $T$. If $\Delta \le \delta$ and $T\ge \bar{T}$, $t^*(\omega_N)\leftarrow T$ and terminate; otherwise, let $T\leftarrow T+1$, $\Delta \leftarrow 0$ and go to Step 2. 
\State Case 2: $\bar{\Omega} \not=\emptyset$ and $\bar{\Omega}\not=\Omega$. Let $\Omega \leftarrow  \Omega \setminus \bar{\Omega}$, $\bar{\Omega}\leftarrow \emptyset$, $T\leftarrow T+1$, $\Delta \leftarrow 0$ and go to Step 2.
\State Case 3: $\bar{\Omega}=\Omega$. Let $\bar{t}(\omega_N)\leftarrow  T$ (if $\bar{t}(\omega_N)=0$) and $t^*(\omega_N)\leftarrow T$ and terminate. 
\end{algorithmic}
\label{algo:nltimeinv}
\end{algorithm}

\section{Invariant set identification}\label{sec:domain}
From the discussions above, we can obtain almost-invariant sets by using a finite set of trajectories with a long horizon. As the sampling number $N$ increases, these sets will become more and more accurate approximations of $O_{\infty}$ with a higher and higher confidence level. We will refer to this step as Phase I. However, the set $O_{t^*(\omega_N)}$ from Algorithm \ref{algo:nltimeinv} is only implicitly defined by the unknown dynamics, which means we will need to take $t^*(\omega_N)$ steps to know whether an initial state is inside $O_{t^*(\omega_N)}$. To circumvent this issue, we will develop a set identification procedure to identify $O_{t^*(\omega_N)}$ explicitly. This will be referred to as Phase II. The overall two-phase procedure is summarized as follows. In Phase I, we use a relatively small set of trajectories with
a long horizon to obtain $t^*(\omega_N)$ and thus $O_{t^*(\omega_N)}$. In Phase II,  we will use a much larger set of trajectories with a short horizon $t^*(\omega_N)$ to identify $O_{t^*(\omega_N)}$ explicitly. For ease of discussion, we assume $f: \mathbb{R}^n \rightarrow \mathbb{R}^n$ is continuous. With this additional assumption, $O_t$ is compact for all $t\in \mathbb{Z}^+$.

Before we present the set identification procedure, we need a reference data set, which will be used to construct an approximation of $O_{t^*(\omega_N)}$. Let the number of initial states in the reference set be denoted by $\bar{N}$ and the reference set be denoted by $\bar{\omega}_{\bar{N}}$. The first step of the set identification procedure is to split the set $\bar{\omega}_{\bar{N}}$ into two disjoint subsets:
\begin{align}
\bar{\omega}^{I}_{\bar{N}}:&=\{x\in \bar{\omega}_{\bar{N}}: x\in O_{t^*(\omega_N)}\}, \label{eqn:omegaNbarI}\\
\bar{\omega}^{O}_{\bar{N}}: & =\{x\in \bar{\omega}_{\bar{N}}: x\not\in O_{t^*(\omega_N)}\}, \label{eqn:omegaNbarO}
\end{align}
which denote the sets of points inside and outside $O_{t^*(\omega_N)}$ respectively.

\subsection{Inner and outer approximations}
With the two disjoint subsets $\bar{\omega}^{I}_{\bar{N}}$ and $\bar{\omega}^{O}_{\bar{N}}$, we will construct inner and outer approximations of $O_{t^*(\omega_N)}$.  For any subsets $S_1,S_2\subseteq \mathbb{R}^n$, any $r\in \mathbb{R}$, and the given set $X$, the following set is defined:
\begin{align}\label{eqn:PiS}
\Pi(S_1,S_2,r):=
\{x\in X: &\inf_{\tilde{x}\in S_1} \|\tilde{x}-x\| \nonumber \\
&-\inf_{\tilde{y}\in S_2}\|\tilde{y}-x\|\le r\}.
\end{align}
We will show that there exists $\delta >0$ such that $\Pi(\bar{\omega}^I_{\bar{N}},\bar{\omega}^O_{\bar{N}},-\delta)$ and $\Pi(\bar{\omega}^I_{\bar{N}},\bar{\omega}^O_{\bar{N}},\delta)$ are inner and outer approximations of $O_{t^*(\omega_N)}$ respectively for the given $\bar{\omega}^I_{\bar{N}}$ and $\bar{\omega}^O_{\bar{N}}$ in (\ref{eqn:omegaNbarI})-(\ref{eqn:omegaNbarO}). Before we establish such results, we will first discuss the properties of the set defined in (\ref{eqn:PiS}) in the following lemma.

\begin{lemma}\label{lem:subset}
Given a compact set $X\subseteq \mathbb{R}^n$, the set defined in (\ref{eqn:PiS}) has the following properties:\\
(i) For any compact subset $S\subseteq X$, $\Pi(S,X\setminus S,0)=S$.\\
(ii) Given any subsets $S_1,S_2,S_1',S_2' \subseteq \mathbb{R}^n$ with $S_1 \subseteq S_1'$ and $S_2 \supseteq S_2'$, we know that $\Pi(S_1,S_2,r)\subseteq\Pi(S'_1,S'_2,r)$ for any $r\in \mathbb{R}$.\\
(iii) For any subsets $S_1,S_2\subseteq \mathbb{R}^n$, $\delta \ge 0$, and $r\in \mathbb{R}$, 
\begin{align}
\Pi(S_1+ \delta \mathcal{B}_n,S_2,r)&\subseteq \Pi(S_1,S_2,r+\delta) \label{eqn:S1S2deltaplus}\\
\Pi(S_1,S_2+ \delta \mathcal{B}_n,r)&\supseteq \Pi(S_1,S_2,r-\delta) \label{eqn:S1S2deltaminus}
\end{align}
\end{lemma}
Proof of Lemma \ref{lem:subset}: From the definition of $\Pi(S,X\setminus S,0)$, we can see that for any $x\in S$, $x\in \Pi(S,X\setminus S,0)$. Hence, $S\subseteq \Pi(S,X\setminus S,0)$. We only need to show that $\Pi(S,X\setminus S,0) \subseteq S$. Suppose there exists a point $x\in \Pi(S,X\setminus S,0)$ but $x \notin S$. This means that $x\in X\setminus S$. Then, we can get 
\begin{align}
\inf_{\tilde{x}\in S} \|\tilde{x}-x\|-\inf_{\tilde{y}\in X\setminus S}\|\tilde{y}-x\| = \inf_{\tilde{x}\in S}\|\tilde{x}-x\| > 0,
\end{align}
which contradicts the fact that $x\in \Pi(S,X\setminus S,0)$. Therefore, it holds true that $S\subseteq X$, $\Pi(S,X\setminus S,0)=S$.\\
(ii) This result follows immediately from the fact that, for any $x\in X$,
\begin{align}
&\inf_{\tilde{x}\in S_1} \|\tilde{x}-x\|-\inf_{\tilde{y}\in S_2}\|\tilde{y}-x\| \nonumber\\
\ge &\inf_{\tilde{x}\in S_1'} \|\tilde{x}-x\|-\inf_{\tilde{y}\in S_2'}\|\tilde{y}-x\|
\end{align}
(iii) Suppose $x\in \Pi(S_1+ \delta \mathcal{B}_n,S_2,r)$, we can see that
\begin{align}
r& \ge \inf_{\tilde{x}\in S_1+ \delta \mathcal{B}_n} \|\tilde{x}-x\|-\inf_{\tilde{y}\in S_2}\|\tilde{y}-x\| \nonumber\\
=&\inf_{\tilde{x}\in S_1, \|z\|\in  \delta} \|\tilde{x}+z-x\|-\inf_{\tilde{y}\in S_2}\|\tilde{y}-x\| \nonumber\\
\ge &\inf_{\tilde{x}\in S_1} \|\tilde{x}-x\|-\inf_{\tilde{y}\in S_2}\|\tilde{y}-x\|-\delta
\end{align}
Hence, $x\in \Pi(S_1,S_2,r+\delta)$, which implies $\Pi(S_1+ \delta \mathcal{B}_n,S_2,r)\subseteq \Pi(S_1,S_2,r+\delta)$. Using similar arguments, we can also show (\ref{eqn:S1S2deltaminus}). For any $x\in \Pi(S_1,S_2,r-\delta)$, we have that
\begin{align}
r-\delta \ge& \inf_{\tilde{x}\in S_1} \|\tilde{x}-x\|-\inf_{\tilde{y}\in S_2}\|\tilde{y}-x\|\nonumber\\
=& \inf_{\tilde{x}\in S_1} \|\tilde{x}-x\|-\inf_{\tilde{y}\in S_2,z\in \delta \mathcal{B}_n}\|\tilde{y}+z-x+z\| \nonumber\\
\ge & \inf_{\tilde{x}\in S_1} \|\tilde{x}-x\|-\inf_{\tilde{y}\in S_2+ \delta \mathcal{B}_n}\|\tilde{y}+z-x\|-\delta \nonumber\\
=& \inf_{\tilde{x}\in S_1} \|\tilde{x}-x\|-\inf_{\tilde{y}\in S_2,z\in \delta \mathcal{B}_n}\|\tilde{y}+z-x\|-\delta
\end{align}
Hence, $x\in \Pi(S_1,S_2+ \delta \mathcal{B}_n,r)$. 
$\Box$

With the properties in Lemma \ref{lem:subset}, we can obtain  inner and outer approximations of $O_{t^*(\omega_N)}$, as stated in the following theorem.

\begin{theorem}\label{thm:delta}
For the given reference set $\bar{\omega}_{\bar{N}}$, let $\bar{\omega}^I_{\bar{N}}$ and $\bar{\omega}^O_{\bar{N}}$ be defined in (\ref{eqn:omegaNbarI}) and (\ref{eqn:omegaNbarO}) respectively. Suppose there exists a $\delta >0$ such that $O_{t^*(\omega_N)} \subseteq \bar{\omega}^I_{\bar{N}}+ \delta \mathcal{B}_n$ and $X\setminus O_{t^*(\omega_N)} \subseteq \bar{\omega}^O_{\bar{N}}+ \delta \mathcal{B}_n$. Then, 
\begin{align}\label{eqn:PiOtPidelta}
\Pi(\bar{\omega}^I_{\bar{N}},\bar{\omega}^O_{\bar{N}},-\delta)\subseteq O_{t^*(\omega_N)} \subseteq \Pi(\bar{\omega}^I_{\bar{N}},\bar{\omega}^O_{\bar{N}},\delta)
\end{align}
\end{theorem}
Proof of Theorem \ref{thm:delta}: We first prove that $  O_{t^*(\omega_N)} \subseteq \Pi(\bar{\omega}^I_{\bar{N}},\bar{\omega}^O_{\bar{N}},\delta)$. From the definition in (\ref{eqn:PiS}), $\bar{\omega}^I_{\bar{N}}+ \delta \mathcal{B}_n \subseteq  \Pi(\bar{\omega}^I_{\bar{N}},\bar{\omega}^O_{\bar{N}},\delta)$, which immediately leads to $  O_{t^*(\omega_N)} \subseteq \Pi(\bar{\omega}^I_{\bar{N}},\bar{\omega}^O_{\bar{N}},\delta)$. Using the properties in Lemma \ref{lem:subset}, we can get have
\begin{align}
O_{t^*(\omega_N)} &= \Pi(O_{t^*(\omega_N)},X\setminus O_{t^*(\omega_N)},0) \nonumber\\
&\supseteq \Pi(O_{t^*(\omega_N)},\bar{\omega}^O_{\bar{N}}+ \delta \mathcal{B}_n,0) \nonumber\\
& \supseteq \Pi(\bar{\omega}^I_{\bar{N}},\bar{\omega}^O_{\bar{N}}+ \delta \mathcal{B}_n,0)  \supseteq \Pi(\bar{\omega}^I_{\bar{N}},\bar{\omega}^O_{\bar{N}},-\delta)
\end{align}
$\Box$

\subsection{Computing the bound}
To get tight approximations, it is desirable to get the minimal $\delta$ such that $O_{t^*(\omega_N)} \subseteq \bar{\omega}^I_{\bar{N}}+ \delta \mathcal{B}_n$ and $X\setminus O_{t^*(\omega_N)} \subseteq \bar{\omega}^O_{\bar{N}}+ \delta \mathcal{B}_n$, where $\bar{\omega}^I_{\bar{N}}$ and $\bar{\omega}^O_{\bar{N}}$ are defined in (\ref{eqn:omegaNbarI}) and (\ref{eqn:omegaNbarO}) for the given reference set $\bar{\omega}_{\bar{N}}$. Let us define
\begin{align}\label{eqn:deltaXmax}
\delta^*(\bar{\omega}_{\bar{N}})  := \max\{&\sup\limits_{x\in O_{t^*(\omega_N)} } \min\limits_{\tilde{x}\in \bar{\omega}^I_{\bar{N}}}\|\tilde{x}-x\|,\nonumber\\
&~\sup\limits_{x\in X\setminus O_{t^*(\omega_N)}}\min\limits_{\tilde{y}\in \bar{\omega}^O_{\bar{N}}}\|\tilde{y}-x\|\}.
\end{align}
For notational simplification, given $O_{t^*(\omega_N)}$ and $\bar{\omega}_{\bar{N}}$, let 
\begin{align}\label{eqn:homegaomega}
h(x):= &\pmb{1}_{O_{t^*(\omega_N)}}(x) h^I(x)  + \pmb{1}_{X\setminus O_{t^*(\omega_N)}}(x) h^O(x). 
\end{align}
where
\begin{align}
h^I(x) := \min\limits_{\tilde{x}\in \bar{\omega}^I_{\bar{N}}}\|\tilde{x}-x\|  \textrm{ and } h^O(x) := \min\limits_{\tilde{y}\in \bar{\omega}^O_{\bar{N}}}\|\tilde{y}-x\|
\end{align}
With the functions above, $\delta^*(\bar{\omega}_{\bar{N}})$ can be equivalently expressed as
\begin{align}\label{eqn:deltaX}
\delta^*(\bar{\omega}_{\bar{N}}) = \sup\limits_{x\in X} h(x).
\end{align}
In the following lemma, we show that $\delta^*(\bar{\omega}_{\bar{N}}) $ defined in (\ref{eqn:deltaXmax}) gives us guaranteed inner and outer approximations.

\begin{lemma}\label{lem:delta}
For the given reference set $\bar{\omega}_{\bar{N}}$ with $\bar{\omega}^I_{\bar{N}}$ and $\bar{\omega}^O_{\bar{N}}$ being defined in (\ref{eqn:omegaNbarI}) and (\ref{eqn:omegaNbarO}) respectively, let  $\delta^*(\bar{\omega}_{\bar{N}})$ be defined in (\ref{eqn:deltaXmax}). Then, we can get that  $O_{t^*(\omega_N)} \subseteq \bar{\omega}^I_{\bar{N}}+ \delta^*(\bar{\omega}_{\bar{N}}) \mathcal{B}_n$ and $X\setminus O_{t^*(\omega_N)} \subseteq \bar{\omega}^O_{\bar{N}}+ \delta^*(\bar{\omega}_{\bar{N}}) \mathcal{B}_n$.
\end{lemma}
Proof of Lemma \ref{lem:delta}: The proof follows directtly from the definition of $\delta^*(\bar{\omega}_{\bar{N}})$ in (\ref{eqn:deltaXmax}). Suppose there exists a point $x\in O_{t^*(\omega_N)}$ but $x\not\in\bar{\omega}^I_{\bar{N}}+ \delta^*(\bar{\omega}_{\bar{N}}) \mathcal{B}_n$, i.e., $\min\limits_{\tilde{x}\in \bar{\omega}^I_{\bar{N}}}\|\tilde{x}-x\|>\delta^*(\bar{\omega}_{\bar{N}})$. This contradicts the fact that $\sup\limits_{x\in O_{t^*(\omega_N)} } \min\limits_{\tilde{x}\in \bar{\omega}^I_{\bar{N}}}\|\tilde{x}-x\| \le \delta^*(\bar{\omega}_{\bar{N}})$. Hence, $O_{t^*(\omega_N)} \subseteq \bar{\omega}^I_{\bar{N}}+ \delta^*(\bar{\omega}_{\bar{N}}) \mathcal{B}_n$. Similarly, we can also show that $X\setminus O_{t^*(\omega_N)} \subseteq \bar{\omega}^O_{\bar{N}}+ \delta^*(\bar{\omega}_{\bar{N}}) \mathcal{B}_n$.
$\Box$

In a Monte-Carlo fashion, $\delta^*(\bar{\omega}_{\bar{N}}) $ will become smaller and smaller as $\bar{N}$ increases. This is in fact a random covering problem \cite{ART:J86}. In general, as shown in \cite{ART:J86}, only asymptotic results can be obtained when $\delta^*(\bar{\omega}_{\bar{N}}) $ goes to $0$. This is usually infeasible in practice. Instead, this paper seeks to numerically approximate $\delta^*(\bar{\omega}_{\bar{N}}) $. Again, we will establish a probabilistic guarantee on the inner and outer approximations from an approximate numerical solution of $\delta^*(\bar{\omega}_{\bar{N}}) $ in the spirit of the scenario approach \cite{ART:CC05,ART:CC06,ART:C10}.

Given $N_{\delta}$ randomly sampled points inside $X$ according to $\mathbb{P}$, denoted by $\tilde{\omega}_{N_{\delta}}$,  the sampled problem of Problem (\ref{eqn:deltaX}) is given by
\begin{subequations}\label{eqn:deltasample}
\begin{align}
\bar{\mathcal{P}}(\tilde{\omega}_{N_{\delta}};\bar{\omega}_{\bar{N}}): ~~&\min_{\delta \ge 0} \delta\\
 \textrm{s.t.} ~~  &h(x) \le \delta, \forall x\in \tilde{\omega}_{N_{\delta}}. \label{eqn:tildexys}
\end{align}
\end{subequations}
where $h(x;\bar{\omega}_{\bar{N}})$ is defined in (\ref{eqn:homegaomega}). Let $\delta^*(\tilde{\omega}_{N_{\delta}};\bar{\omega}_{\bar{N}})$ denote the solution of the sampled problem above. We define the violation probability of $\bar{\mathcal{P}}(\tilde{\omega}_{N_{\delta}};\bar{\omega}_{\bar{N}})$ by
\begin{align}\label{eqn:Vtilde}
\tilde{V}(\tilde{\omega}_{N_{\delta}}) =\mathbb{P}\{x\in X: h(x)> \delta^*(\tilde{\omega}_{N_{\delta}};\bar{\omega}_{\bar{N}}) \}
\end{align}
Let
\begin{align}
\tilde{V}^I(\tilde{\omega}_{N_{\delta}})  &= \mathbb{P}(x\in O_{t^*(\omega_N)}: h^I(x)>\delta^*(\tilde{\omega}_{N_{\delta}};\bar{\omega}_{\bar{N}})) \label{eqn:VtildeI}\\
\tilde{V}^O(\tilde{\omega}_{N_{\delta}})  &= \mathbb{P}(x\in X\setminus O_{t^*(\omega_N)}: h^O(x)>\delta^*(\tilde{\omega}_{N_{\delta}};\bar{\omega}_{\bar{N}})). \label{eqn:VtildeO}
\end{align}
From (\ref{eqn:homegaomega}), we can see that
\begin{align}\label{eqn:VIVO}
\tilde{V}(\tilde{\omega}_{N_{\delta}}) =\tilde{V}^I(\tilde{\omega}_{N_{\delta}})  + \tilde{V}^O(\tilde{\omega}_{N_{\delta}})   
\end{align}
In fact, $\tilde{V}^I(\tilde{\omega}_{N_{\delta}})$ is the measure of the violating subset in $O_{t^*(\omega_N)}$ and $\tilde{V}^O(\tilde{\omega}_{N_{\delta}})$ is the measure of the violating subset in $X\setminus O_{t^*(\omega_N)}$. Adapted from Theorem 3.3 in \cite{ART:C10}, the following theorem can be obtained.
\begin{theorem}\label{thm:deltarandom}
Consider a given reference data set $\bar{\omega}_{\bar{N}}$ with $\bar{\omega}^I_{\bar{N}}$ and $\bar{\omega}^O_{\bar{N}}$ being defined in (\ref{eqn:omegaNbarI}) and (\ref{eqn:omegaNbarO}) respectively. Let a set of $N_{\delta}$ points, denoted by $\tilde{\omega}_{N_{\delta}}$, be randomly extracted according to $\mathbb{P}$ over $X$. Let $\delta^*(\tilde{\omega}_{N_{\delta}};\bar{\omega}_{\bar{N}})$ be obtained by solving $\bar{\mathcal{P}}(\tilde{\omega}_{N_{\delta}};\bar{\omega}_{\bar{N}})$. For any $\tilde{\epsilon}>0$, let $\beta_{\delta} = (1-\tilde{\epsilon})^{N_{\delta}}$. Then, it holds that
%\begin{align}
%&\mathbb{P}\{O_{t^*(\omega_N)} \cap \Pi(\bar{\omega}^I_{\bar{N}},\bar{\omega}^O_{\bar{N}},\delta^*(\tilde{\omega}_{N_{\delta}};\bar{\omega}_{\bar{N}})) \}  \nonumber\\
%&\ge \mathbb{P}\{O_{t^*(\omega_N)}\}-\tilde{\epsilon} \label{eqn:epsilontouter}\\
%&\mathbb{P}\{O_{t^*(\omega_N)} \cup \Pi(\bar{\omega}^I_{\bar{N}},\bar{\omega}^O_{\bar{N}},-\delta^*(\tilde{\omega}_{N_{\delta}};\bar{\omega}_{\bar{N}})) \}\nonumber \\
%&\le \mathbb{P}\{O_{t^*(\omega_N)}\} +\tilde{\epsilon} \label{eqn:epsilontinner}
%\end{align}
\begin{align}
\mathbb{P}^{N_{\delta}}\{\tilde{\omega}_{N_{\delta}} \in X^{N_{\delta}}:&\mathbb{P}\{\underline{\chi}\}-\tilde{\epsilon} \label{eqn:epsilonbounds}\\
&\le \mathbb{P}\{O_{t^*(\omega_N)}\} \le \mathbb{P}\{\overline{\chi} \} + \tilde{\epsilon} \} \ge 1-\beta_{\delta}. \nonumber
\end{align}
where
\begin{align}
\overline{\chi} & : = O_{t^*(\omega_N)} \cap \Pi(\bar{\omega}^I_{\bar{N}},\bar{\omega}^O_{\bar{N}},\delta^*(\tilde{\omega}_{N_{\delta}};\bar{\omega}_{\bar{N}}))\\
\underline{\chi} & : = O_{t^*(\omega_N)} \cup \Pi(\bar{\omega}^I_{\bar{N}},\bar{\omega}^O_{\bar{N}},-\delta^*(\tilde{\omega}_{N_{\delta}};\bar{\omega}_{\bar{N}}))
\end{align}
\end{theorem}
Proof of Theorem \ref{thm:deltarandom}: We will use the notation $\tilde{V}(\tilde{\omega}_{N_{\delta}}), \tilde{V}^I(\tilde{\omega}_{N_{\delta}}),$ and $ \tilde{V}^O(\tilde{\omega}_{N_{\delta}})$ in (\ref{eqn:Vtilde})-(\ref{eqn:VtildeO}) in the proof. From Theorem 3.3 in \cite{ART:C10}, we know that
\begin{align}
\mathbb{P}^{N_{\delta}}\{\tilde{\omega}_{N_{\delta}} \in X^{N_{\delta}}: \tilde{V}(\tilde{\omega}_{N_{\delta}})   >\tilde{\epsilon}\} \le \beta_{\delta}.
\end{align}
Hence, to show (\ref{eqn:epsilonbounds}), we only need to show that $\tilde{V}(\tilde{\omega}_{N_{\delta}})   \le \tilde{\epsilon}$ implies $\mathbb{P}\{\underline{\chi}\}-\tilde{\epsilon} \le \mathbb{P}\{O_{t^*(\omega_N)}\} \le \mathbb{P}\{\overline{\chi} \} + \tilde{\epsilon}$. First, we show that $\mathbb{P}\{O_{t^*(\omega_N)}\} \le \mathbb{P}\{\overline{\chi} \} + \tilde{\epsilon}$. From (\ref{eqn:VIVO}), we can see that
$
\tilde{V}^I(\tilde{\omega}_{N_{\delta}}) \le \tilde{\epsilon},
$
which lead to 
\begin{align}\label{eqn:POxnotinB}
\mathbb{P}\{x\in O_{t^*(\omega_N)}:x\notin \bar{\omega}^I_{\bar{N}}+ \delta^*(\tilde{\omega}_{N_{\delta}};\bar{\omega}_{\bar{N}}) \mathcal{B}_n\} \le \tilde{\epsilon}
\end{align}
since 
$
\{x\in O_{t^*(\omega_N)}: h^I(x)>\delta^*(\tilde{\omega}_{N_{\delta}};\bar{\omega}_{\bar{N}})\} = \{x\in O_{t^*(\omega_N)}:x\notin \bar{\omega}^I_{\bar{N}}+ \delta^*(\tilde{\omega}_{N_{\delta}};\bar{\omega}_{\bar{N}}) \mathcal{B}_n\}
$
from the same arguments in the proof of Lemma \ref{lem:delta}. From the definition in (\ref{eqn:PiS}), we know that $\bar{\omega}^I_{\bar{N}}+ \delta^*(\tilde{\omega}_{N_{\delta}};\bar{\omega}_{\bar{N}}) \mathcal{B}_n \subseteq \Pi(\bar{\omega}^I_{\bar{N}},\bar{\omega}^O_{\bar{N}},\delta^*(\tilde{\omega}_{N_{\delta}};\bar{\omega}_{\bar{N}}))$. Hence, from  (\ref{eqn:POxnotinB}), we can get $\mathbb{P}\{O_{t^*(\omega_N)} \setminus \Pi(\bar{\omega}^I_{\bar{N}},\bar{\omega}^O_{\bar{N}},\delta^*(\tilde{\omega}_{N_{\delta}};\bar{\omega}_{\bar{N}}))\}\le \tilde{\epsilon}$, which implies that $\mathbb{P}\{O_{t^*(\omega_N)}\} \le \mathbb{P}\{\overline{\chi} \} + \tilde{\epsilon}$. Now, we will show that $\mathbb{P}\{\underline{\chi}\}-\tilde{\epsilon} \le \mathbb{P}\{O_{t^*(\omega_N)}\}$. Again, from (\ref{eqn:VIVO}), we can see that
$
\tilde{V}^O(\tilde{\omega}_{N_{\delta}}) \le \tilde{\epsilon},
$
which lead to 
\begin{align}\label{eqn:POxnotinBout}
\mathbb{P}\{x\in X\setminus O_{t^*(\omega_N)}:x\notin \bar{\omega}^O_{\bar{N}}+ \delta^*(\tilde{\omega}_{N_{\delta}};\bar{\omega}_{\bar{N}}) \mathcal{B}_n\} \le \tilde{\epsilon}.
\end{align}
It can be shown that
%$
%(X\setminus O_{t^*(\omega_N)}) \cap \bar{\omega}^O_{\bar{N}}+ \delta^*(\tilde{\omega}_{N_{\delta}};\bar{\omega}_{\bar{N}}) \mathcal{B}_n\} \subseteq \bar{\omega}^O_{\bar{N}}+ \delta^*(\tilde{\omega}_{N_{\delta}};\bar{\omega}_{\bar{N}}) \mathcal{B}_n
%$
$
\bar{\omega}^O_{\bar{N}}+ \delta^*(\tilde{\omega}_{N_{\delta}};\bar{\omega}_{\bar{N}}) \mathcal{B}_n \supseteq (X\setminus O_{t^*(\omega_N)}) \cap \bar{\omega}^O_{\bar{N}}+ \delta^*(\tilde{\omega}_{N_{\delta}};\bar{\omega}_{\bar{N}}) \mathcal{B}_n\} 
$
and that 
$$
\bar{\omega}^I_{\bar{N}} \subseteq O_{t^*(\omega_N)} \subseteq X \setminus\left((X\setminus O_{t^*(\omega_N)})\cap \bar{\omega}^O_{\bar{N}}+ \delta^*(\tilde{\omega}_{N_{\delta}};\bar{\omega}_{\bar{N}}) \mathcal{B}_n\} \right) .
$$
Using the properties in Lemma \ref{lem:subset}, we can get
\begin{align}
&\mathbb{P}\{X \setminus\left((X\setminus O_{t^*(\omega_N)})\cap \bar{\omega}^O_{\bar{N}}+ \delta^*(\tilde{\omega}_{N_{\delta}};\bar{\omega}_{\bar{N}}) \mathcal{B}_n\} \right) \} \nonumber\\
=& \mathbb{P}\{\Pi(X \setminus\left((X\setminus O_{t^*(\omega_N)})\cap \bar{\omega}^O_{\bar{N}}+ \delta^*(\tilde{\omega}_{N_{\delta}};\bar{\omega}_{\bar{N}}) \mathcal{B}_n\} \right), \nonumber\\
&~~(X\setminus O_{t^*(\omega_N)})\cap \bar{\omega}^O_{\bar{N}}+ \delta^*(\tilde{\omega}_{N_{\delta}};\bar{\omega}_{\bar{N}}) \mathcal{B}_n\},0)\} \nonumber\\
\ge & \mathbb{P}\{\Pi(\bar{\omega}^I_{\bar{N}},\bar{\omega}^O_{\bar{N}}+ \delta^*(\tilde{\omega}_{N_{\delta}};\bar{\omega}_{\bar{N}}) \mathcal{B}_n,0)\cup O_{t^*(\omega_N)}\}\nonumber \\
\ge & \mathbb{P}\{\Pi(\bar{\omega}^I_{\bar{N}},\bar{\omega}^O_{\bar{N}},-\delta^*(\tilde{\omega}_{N_{\delta}};\bar{\omega}_{\bar{N}})\cup O_{t^*(\omega_N)}\} \label{eqn:XminusdeltaminusP}
\end{align}
From (\ref{eqn:POxnotinBout}) and (\ref{eqn:XminusdeltaminusP}), we know that
\begin{align}
\tilde{\epsilon} \ge& \mathbb{P}\{x\in X\setminus O_{t^*(\omega_N)}:x\notin \bar{\omega}^O_{\bar{N}}+ \delta^*(\tilde{\omega}_{N_{\delta}};\bar{\omega}_{\bar{N}}) \mathcal{B}_n\}\\
= & \mathbb{P}\{X\setminus O_{t^*(\omega_N)}\}  \nonumber \\
& -\mathbb{P}\{(X\setminus O_{t^*(\omega_N)}) \cap \bar{\omega}^O_{\bar{N}}+ \delta^*(\tilde{\omega}_{N_{\delta}};\bar{\omega}_{\bar{N}}) \mathcal{B}_n\}\\
=& 1- \mathbb{P}\{O_{t^*(\omega_N)}\} -1 \nonumber\\
 &+ \mathbb{P}\{X \setminus\left((X\setminus O_{t^*(\omega_N)})\cap \bar{\omega}^O_{\bar{N}}+ \delta^*(\tilde{\omega}_{N_{\delta}};\bar{\omega}_{\bar{N}}) \mathcal{B}_n\} \right) \}\nonumber\\
\ge & \mathbb{P}\{\Pi(\bar{\omega}^I_{\bar{N}},\bar{\omega}^O_{\bar{N}},-\delta^*(\tilde{\omega}_{N_{\delta}};\bar{\omega}_{\bar{N}}))\cup O_{t^*(\omega_N)}\} \nonumber\\
&-\mathbb{P}\{O_{t^*(\omega_N)}\}\nonumber
\end{align}
This completes the proof. $\Box$

From Theorem \ref{thm:deltarandom}, we can see that, with probability no smaller than  $1-\beta_{\delta}$, the set $O_{t^*(\omega_N)}$ obtained from Phase I is almost contained in the set $\Pi(\bar{\omega}^I_{\bar{N}},\bar{\omega}^O_{\bar{N}},\delta^*(\tilde{\omega}_{N_{\delta}};\bar{\omega}_{\bar{N}}))$ except a small subset whose measure is bounded by $\tilde{\epsilon}$. Hence, $\Pi(\bar{\omega}^I_{\bar{N}},\bar{\omega}^O_{\bar{N}},\delta^*(\tilde{\omega}_{N_{\delta}};\bar{\omega}_{\bar{N}}))$ can be considered as an almost outer approximation of $O_{t^*(\omega_N)}$ with probability no smaller than  $1-\beta_{\delta}$. Similarly, we can also claim that $\Pi(\bar{\omega}^I_{\bar{N}},\bar{\omega}^O_{\bar{N}},-\delta^*(\tilde{\omega}_{N_{\delta}};\bar{\omega}_{\bar{N}}))$ an almost inner approximation of $O_{t^*(\omega_N)}$ with probability no smaller than  $1-\beta_{\delta}$.

\subsection{An iterative set identification procedure}
For the given $\tilde{\epsilon}$ and $\tilde{\beta}$, we can obtain $N_{\delta}$ such that $\beta_{\delta}$ defined in Theorem \ref{thm:deltarandom} is smaller than $\tilde{\beta}$. We will then sample the set $\tilde{\omega}_{N_{\delta}}$ with $N_{\delta}$ points and solve $\bar{\mathcal{P}}(\tilde{\omega}_{N_{\delta}};\bar{\omega}_{\bar{N}})$. If the solution $\delta^*(\tilde{\omega}_{N_{\delta}};\bar{\omega}_{\bar{N}})$ is less than some targeted value $\bar{\delta}$, we will consider that the reference set $\bar{\omega}_{\bar{N}}$ has enough points and the procedure is terminated. Otherwise,  the points in $\tilde{\omega}_{N_{\delta}}$ will be added to the reference set $\bar{\omega}_{\bar{N}}$ and the procedure continues. The set $\tilde{\omega}_{N_{\delta}}$ can be considered as a test set for the current set $\bar{\omega}_{\bar{N}}$. The set identification procedure is formally described in Algorithm \ref{algo:domain}.

\begin{algorithm}[h]
\caption{The data-driven set identification procedure (Phase II)}
\begin{algorithmic}[1]
\renewcommand{\algorithmicrequire}{\textbf{Input:}}
\renewcommand{\algorithmicensure}{\textbf{Output:}}
\Require $X$, $\mathbb{P}$, $t^*(\omega_N)$, $\omega_N$, $N$, $\tilde{\epsilon}$, $\tilde{\beta}$, $\bar{\delta}$
\Ensure  $\bar{\omega}_{\bar{N}}$, $\delta^*(\tilde{\omega}_{N_{\delta}};\bar{\omega}_{\bar{N}})$
\State  Initialization: Compute $N_{\delta}$ with $\beta_{\delta}<\tilde{\beta}$, let $\bar{N}\leftarrow N$ and $\bar{\omega}_{\bar{N}} \leftarrow \omega_N$, and split the set $\bar{\omega}_{\bar{N}}$ into $\bar{\omega}^I_{\bar{N}}$ and $\bar{\omega}^O_{\bar{N}}$;
\State  Randomly sample $N_{\delta}$ initial states $\tilde{\omega}_{N_{\delta}}$ over $X$ according to $\mathbb{P}$, and generate their trajectories for $t^*(\omega_N)$ steps;
\State Solve $\bar{\mathcal{P}}(\tilde{\omega}_{N_{\delta}};\bar{\omega}_{\bar{N}})$, obtain the solution $\delta^*(\tilde{\omega}_{N_{\delta}};\bar{\omega}_{\bar{N}})$, and let $\bar{\omega}_{\bar{N}}\leftarrow \bar{\omega}_{\bar{N}} \cup \tilde{\omega}_{N_{\delta}}$ and $\bar{N} \leftarrow \bar{N}+N_{\delta}$;
\State If $\delta^*(\tilde{\omega}_{N_{\delta}};\bar{\omega}_{\bar{N}})\le\bar{\delta}$, terminate; otherwise, go to Step 2.
\end{algorithmic}
\label{algo:domain}
\end{algorithm}

\section{Numerical examples}\label{sec:num}
\begin{figure}[h]
  \centering
  \begin{tabular}{cc}
  \subcaptionbox{ \label{fig:nl0p02}}{\includegraphics[width=1.6in]{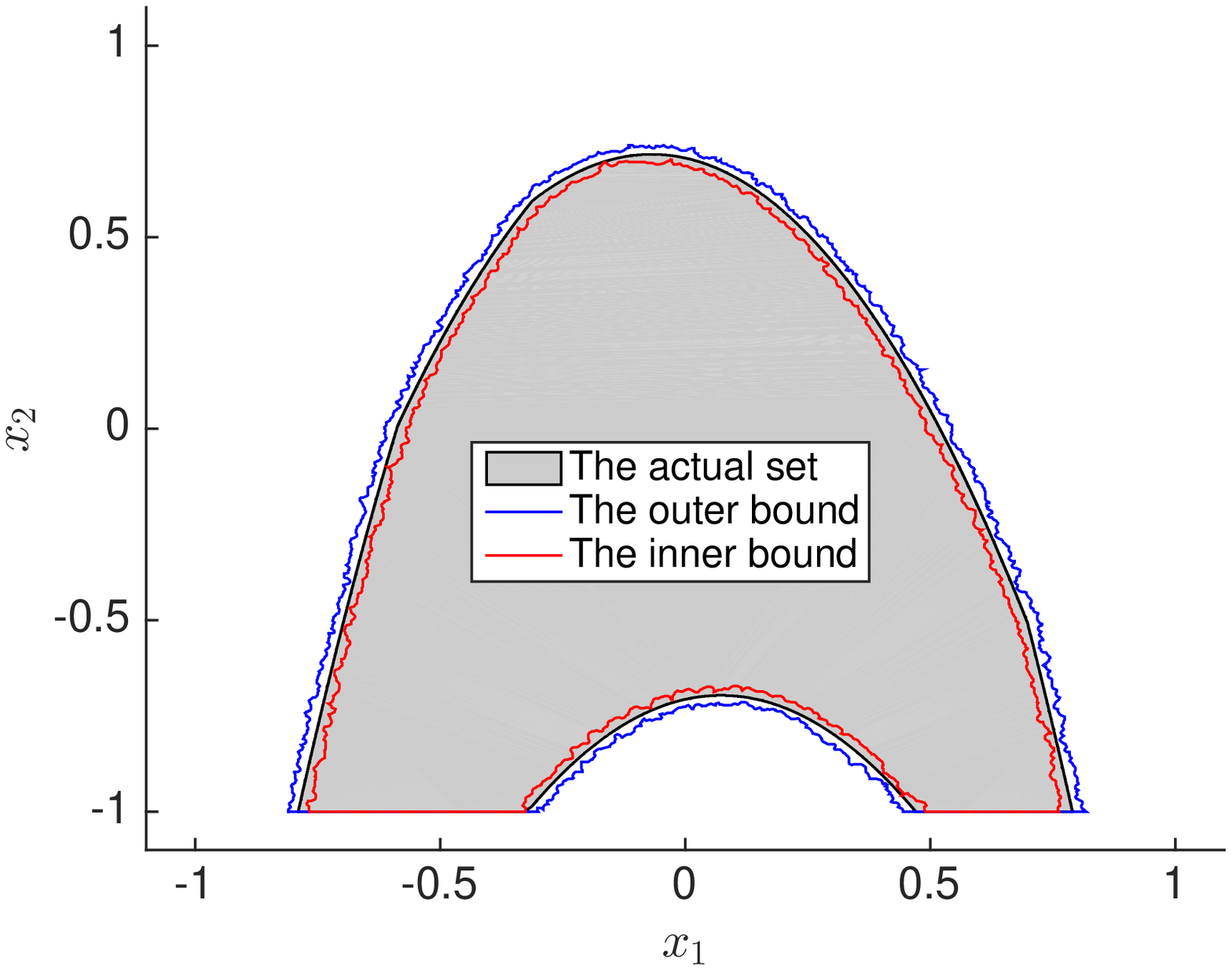}} &
   \subcaptionbox{ \label{fig:nl0p01}}{\includegraphics[width=1.6in]{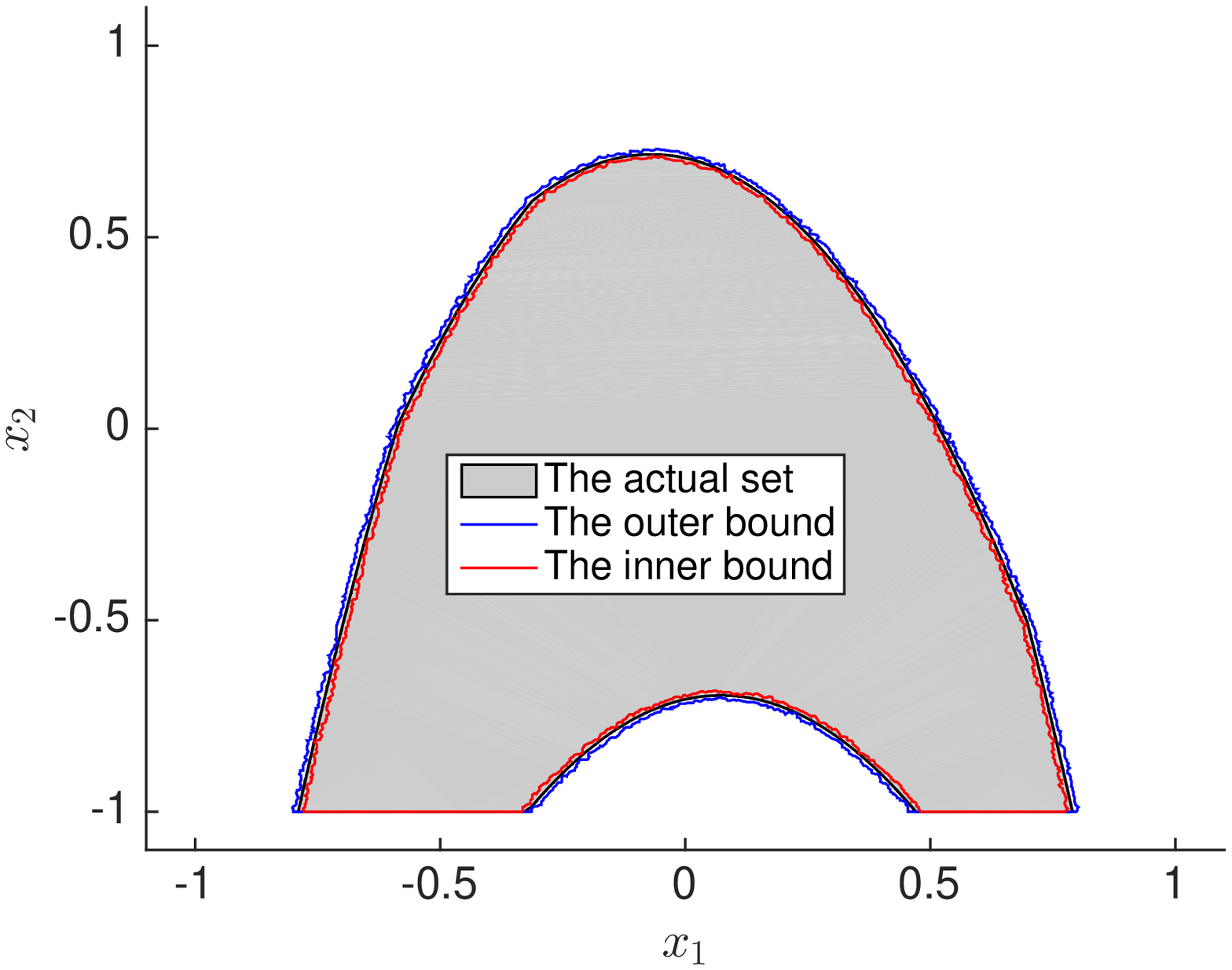}}\\
  \end{tabular}
\caption{Set identification for two different values of $\bar{\delta}$ in Example \ref{exam:nl}: (a) $\bar{\delta}=0.02$ and (b) $\bar{\delta}=0.01$. The blue curve refers to $\Pi(\bar{\omega}^I_{\bar{N}},\bar{\omega}^O_{\bar{N}},\delta^*(\tilde{\omega}_{N_{\delta}};\bar{\omega}_{\bar{N}}))$ (the outer bound) and the red curve refers to $\Pi(\bar{\omega}^I_{\bar{N}},\bar{\omega}^O_{\bar{N}},-\delta^*(\tilde{\omega}_{N_{\delta}};\bar{\omega}_{\bar{N}}))$ (the inner bound). The actual set is computed by gridding.}
\label{fig:nl}
\end{figure}

\begin{figure}[h]
\centering
\includegraphics[width=2.5in]{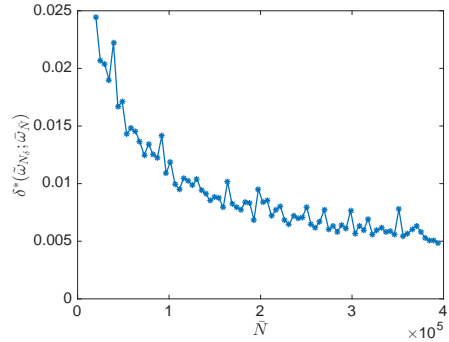}
\caption{The curve of $\delta^*(\tilde{\omega}_{N_{\delta}};\bar{\omega}_{\bar{N}})$ in Example \ref{exam:nl}}
\label{fig:delta}
\end{figure}

In this section, we will consider several examples to illustrate the proposed data-driven framework for computing almost-invariant sets. 
\begin{example}\label{exam:nl}
We first consider the following nonlinear system
\begin{align*}
x_1(t+1) &= 2(x_1(t))^2+x_2(t),\\
x_2(t+1) &= -2\left(2(x_1(t))^2+x_2(t)\right)^2-0.8x_1(t),
\end{align*}
\end{example}
This nonlinear system is asymptotically stable in $\mathbb{R}^2$. The state constraint set is given by $X:=\{x\in \R^2: |x_1|\le 1, |x_2|\le 1\}$. Let $\beta=0.05$ and $\epsilon=10^{-3}$. 
Let us generate $\omega_N$ according to the uniform distribution over $X$ with $N=\lceil \frac{\ln(\beta)}{\ln(1-\epsilon)} \rceil =2995$. The solution of $\mathcal{P}^*(\omega_N)$ is $t^*(\omega_N)=3$. Then, we will use Algorithm \ref{algo:domain} to identify $O_{3}$. Let $\tilde{\epsilon}=10^{-3}$ and $\tilde{\beta}=0.01$. To make sure that $\beta_{\delta}$ in Theorem \ref{thm:deltarandom} is smaller than  $\tilde{\beta}$, we choose $N_{\delta}$ to be $4.8\times 10^3$, which gives us a $\beta_{\delta}$ of value $0.0476$. Two different values $\bar{\delta}$ are selected and the corresponding sets $\Pi(\bar{\omega}^I_{\bar{N}},\bar{\omega}^O_{\bar{N}},\delta^*(\tilde{\omega}_{N_{\delta}};\bar{\omega}_{\bar{N}}))$ and $\Pi(\bar{\omega}^I_{\bar{N}},\bar{\omega}^O_{\bar{N}},-\delta^*(\tilde{\omega}_{N_{\delta}};\bar{\omega}_{\bar{N}}))$ are presented in Figure \ref{fig:nl}. As we can see the trend from Figure \ref{fig:delta}, $\delta^*(\tilde{\omega}_{N_{\delta}};\bar{\omega}_{\bar{N}})$ will eventually decrease as $\bar{N}$ increases.

In the rest of this section, we will show a few examples with more complex dynamics. The existence of their maximal invariant sets is verified numerically by gridding.

\begin{figure}[h]
\centering
\includegraphics[width=2.5in]{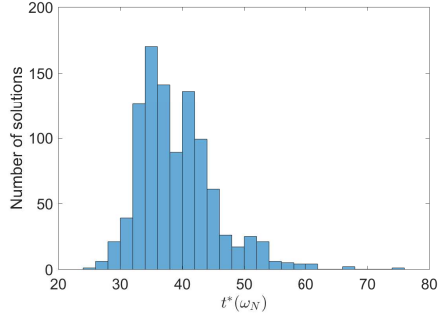}
\caption{The histogram of the distribution of $t^*(\omega_N)$ in Example \ref{exam:lure}.}
\label{fig:histlure}
\end{figure}

\begin{figure}[h]
  \centering
  \begin{tabular}{cc}
  \subcaptionbox{ \label{fig:lure0p3}}{\includegraphics[width=1.6in]{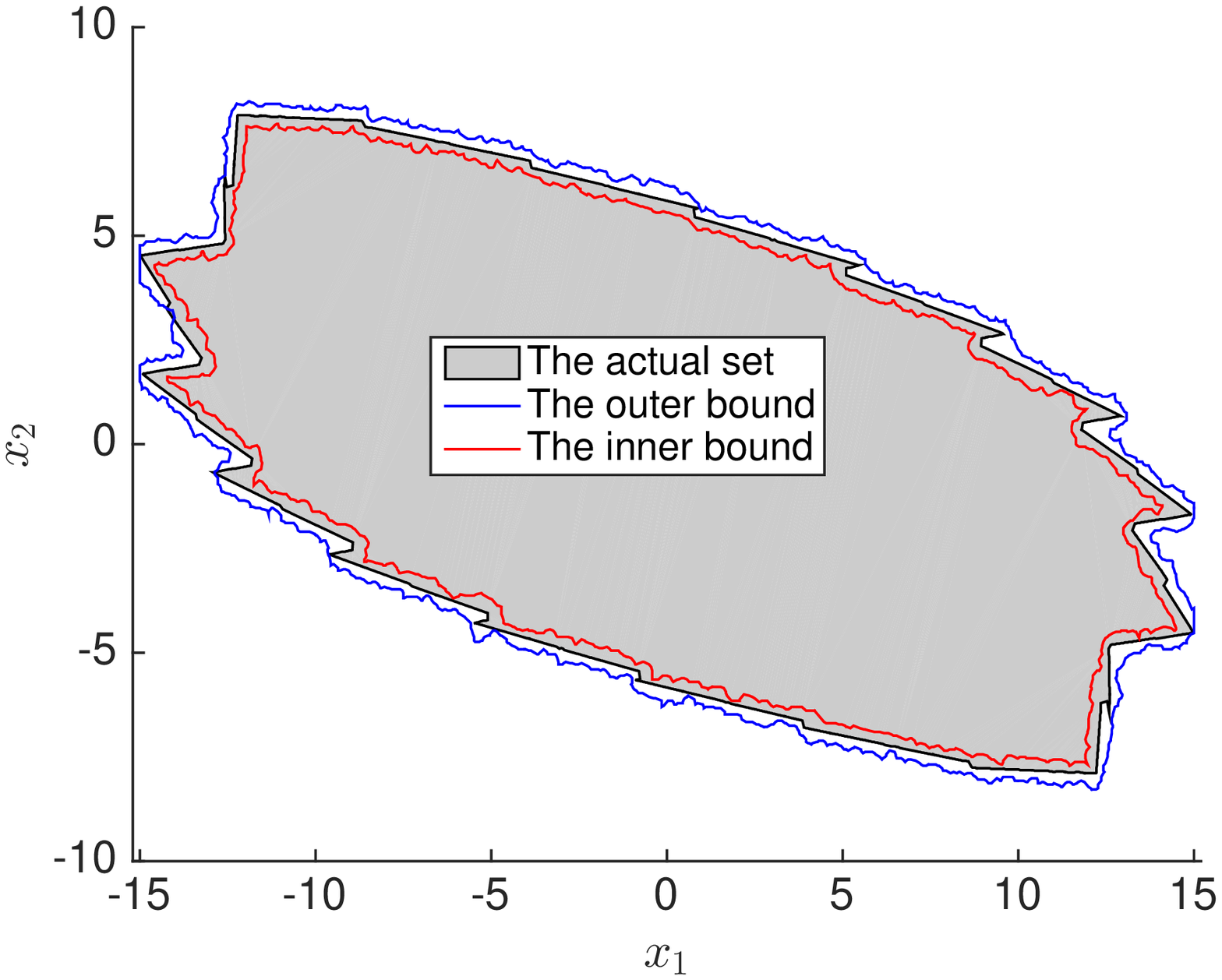}} &
   \subcaptionbox{ \label{fig:lur0p2}}{\includegraphics[width=1.6in]{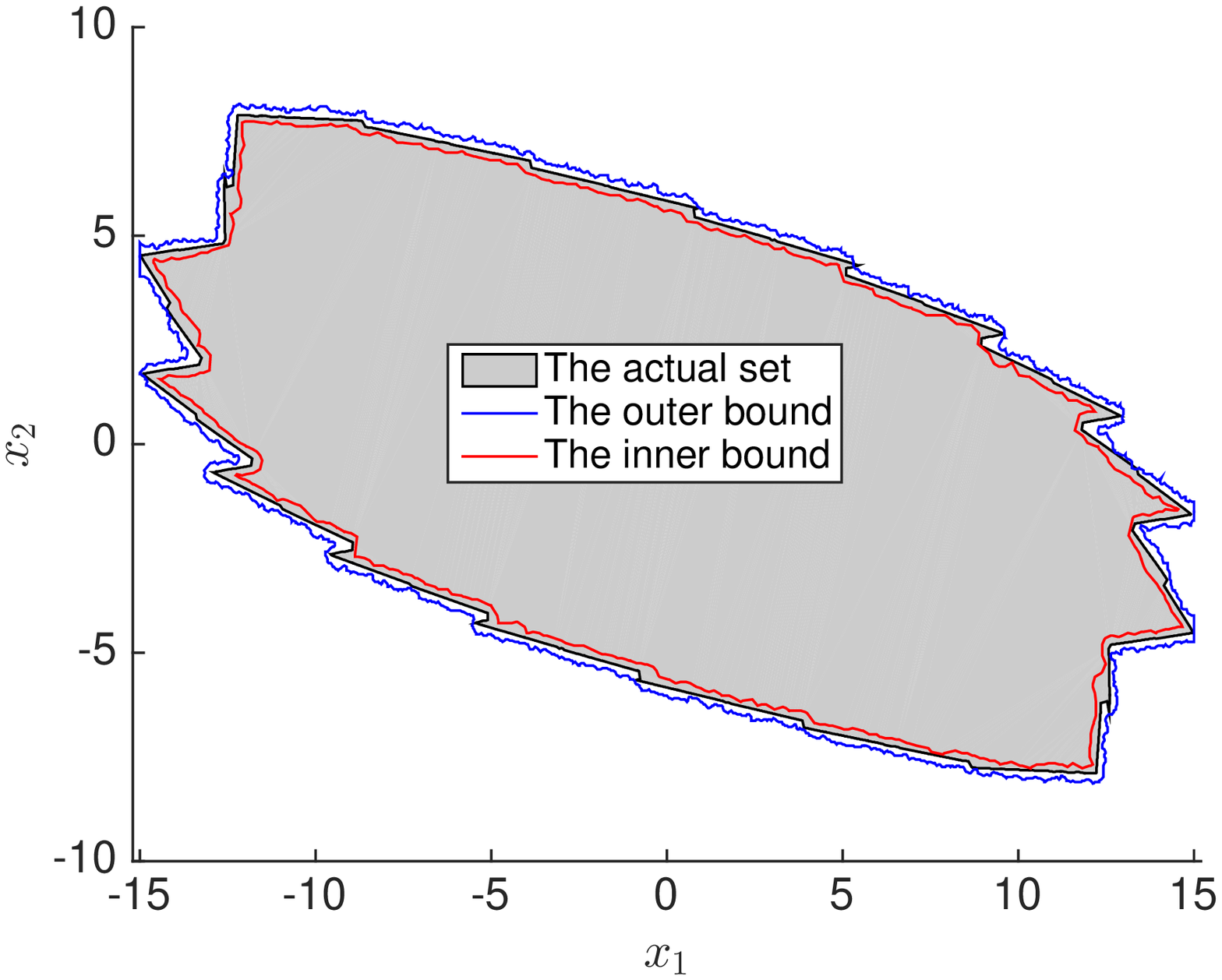}}\\
  \end{tabular}
\caption{Set identification for two different values of $\bar{\delta}$ in Example \ref{exam:lure}: (a) $\bar{\delta}=0.3$ and (b) $\bar{\delta}=0.2$. }
\label{fig:lure}
\end{figure}

\begin{example}\label{exam:lure}
Consider the discrete-time Lur'e system \cite{ART:ACFC09}
\begin{align*}
x(t+1) = Ax(t)-B\phi(F^Tx(t))
\end{align*}
with 
\begin{align*}
A = \begin{bmatrix}
1.2 & 1\\
0 & 1.2
\end{bmatrix}, B = \begin{bmatrix}
0.5 \\
1
\end{bmatrix},F= \begin{bmatrix}
0.6290 \\
1.2261
\end{bmatrix}
\end{align*}
and the function $\phi(y):$
\begin{align*}
\phi(y) &=  \begin{cases} 
      y, & \textrm{if } y\in [0,2)  \\
      0.25y+1.5, & \textrm{if } y\in [2,4) \\
      2.5, & \textrm{if } y\in [4,\infty)
   \end{cases}, \\
\phi(y) &= -\phi(-y), \forall y \le 0
\end{align*}
\end{example}
The state constraint set is given by $X:=\{x\in \R^2: |x_1|\le 15, |x_2|\le 10\}$.  It can be verified that the system is not asymptotically stable in the whole $X$. However, there exists an attractor inside $X$. By gridding, we can see that $O_k$ changes little after $k\ge 20$. We will use the same setting in Example \ref{exam:nl}. Since the solution of $\mathcal{P}^*(\omega_N)$ can be different for different realizations of $\omega_N$, we take $1000$ realizations and the histogram of the solutions is given in Figure \ref{fig:histlure}. As we can see, the solution $t^*(\omega_N)$ varies from $25$ to $76$. Here, we present the result when $t^*(\omega_N)=45$. The inner and outer approximations are given in Figure \ref{fig:lure} for two different values of $\bar{\delta}$.

\begin{figure}[h]
\centering
\includegraphics[width=2.5in]{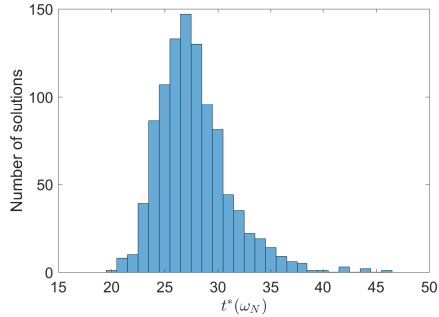}
\caption{The histogram of the distribution of $t^*(\omega_N)$ in Example \ref{exam:chatala}.}
\label{fig:histchatala}
\end{figure}

\begin{figure}[h]
  \centering
  \begin{tabular}{cc}
  \subcaptionbox{ \label{fig:chatala0p05}}{\includegraphics[width=1.6in]{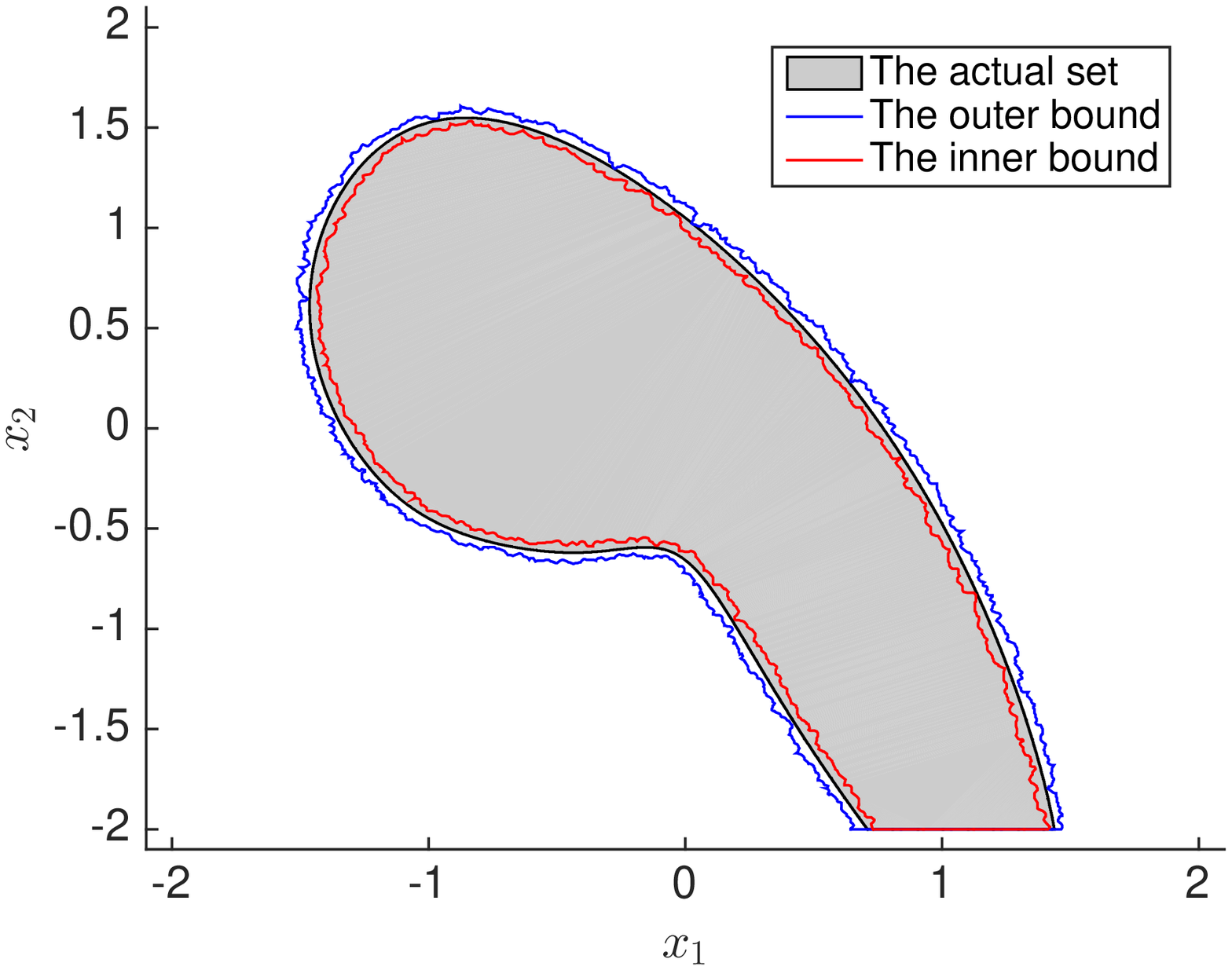}} &
   \subcaptionbox{ \label{fig:chatala0p03}}{\includegraphics[width=1.6in]{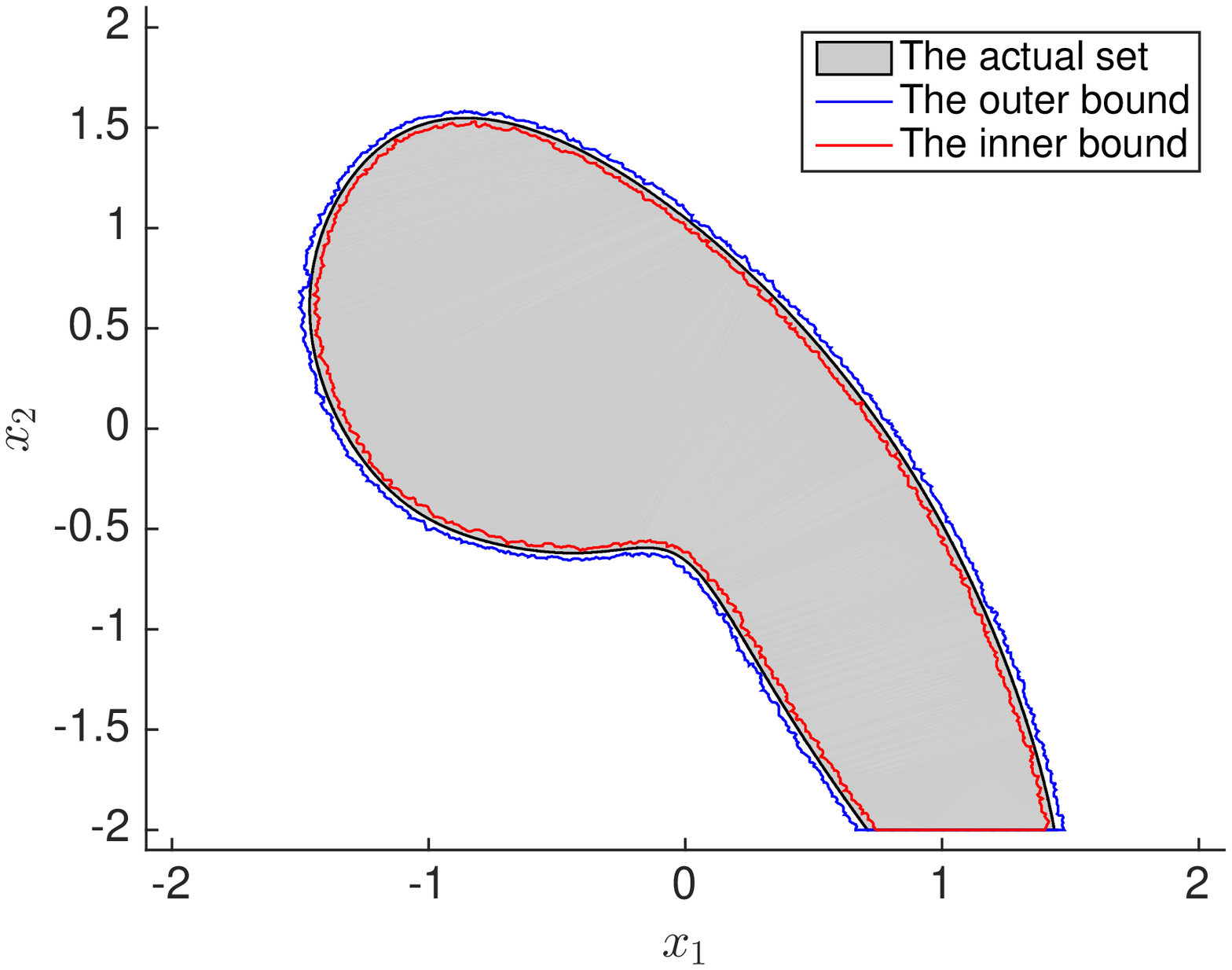}}\\
  \end{tabular}
\caption{Set identification for two different values of $\bar{\delta}$ in Example \ref{exam:chatala}: (a) $\bar{\delta}=0.05$ and (b) $\bar{\delta}=0.03$. }
\label{fig:chatala}
\end{figure}

\begin{example}\label{exam:chatala}
Consider the Chatala system \cite{ART:KK11,ART:KHJ14}
\begin{align*}
x_1(t+1) &= x_1(t)+x_2(t),\\
x_2(t+1) &= -0.5952+(x_1(t))^2,
\end{align*}
\end{example}
The state constraint set is given by $X:=\{x\in \R^2: |x_1|\le 2, |x_2|\le 2\}$. This system is not asymptotically stable at the origin, however, as shown in \cite{ART:KK11,ART:KHJ14}, there is an attractor contained in $X$. Again, we use the same setting in Example \ref{exam:nl} and take $1000$ realizations of $\omega_N$. The solution of $\mathcal{P}^*(\omega_N)$ varies from $20$ to $46$ and the histogram of the solutions is shown in Figure \ref{fig:histchatala}. We plot out the case when $t^*(\omega_N)=31$. The inner and outer approximations are presented in Figure \ref{fig:chatala} for two different values of $\bar{\delta}$

\begin{figure}[h]
  \centering
  \begin{tabular}{cc}
  \subcaptionbox{ \label{fig:pwa0p1}}{\includegraphics[width=1.6in]{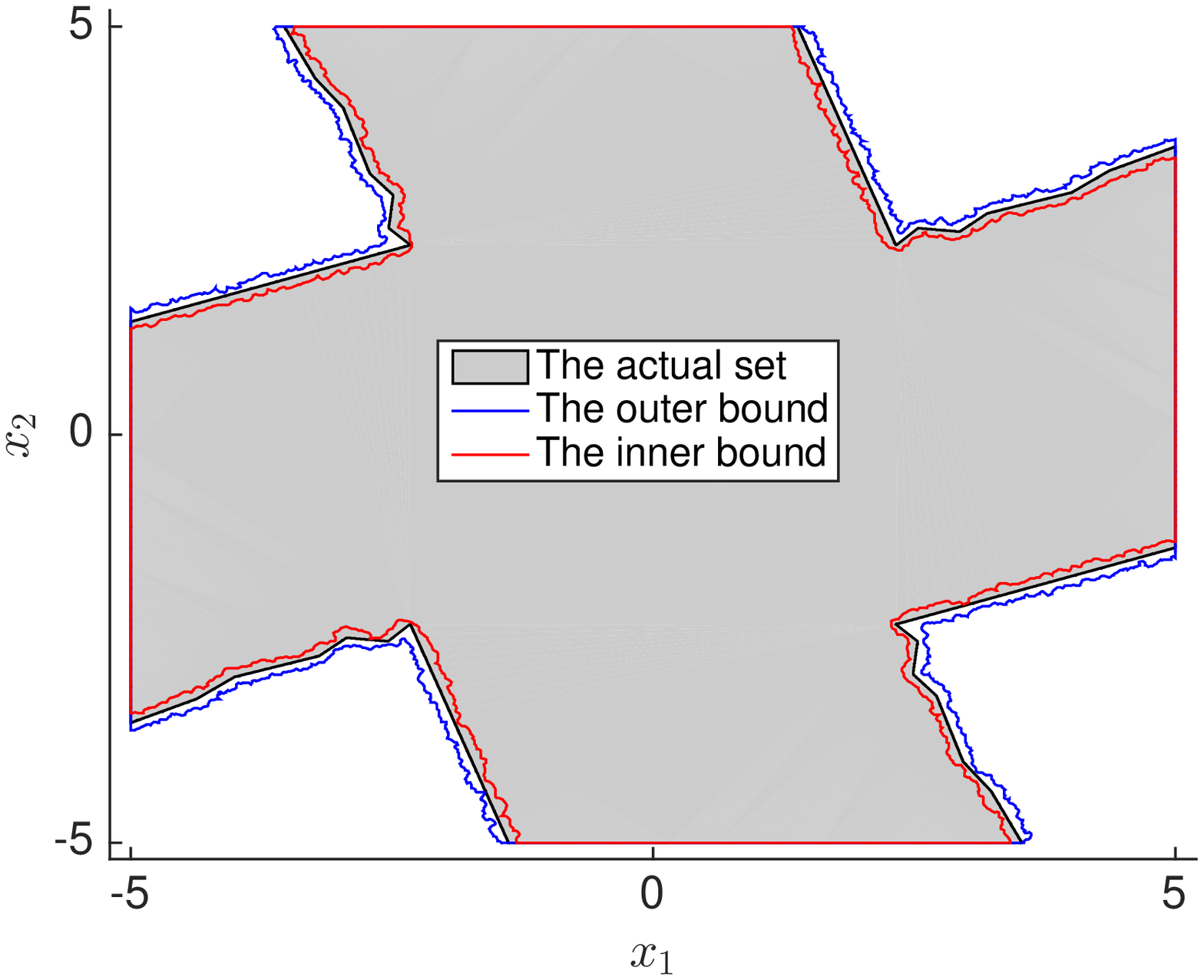}} &
   \subcaptionbox{ \label{fig:pwa0p05}}{\includegraphics[width=1.6in]{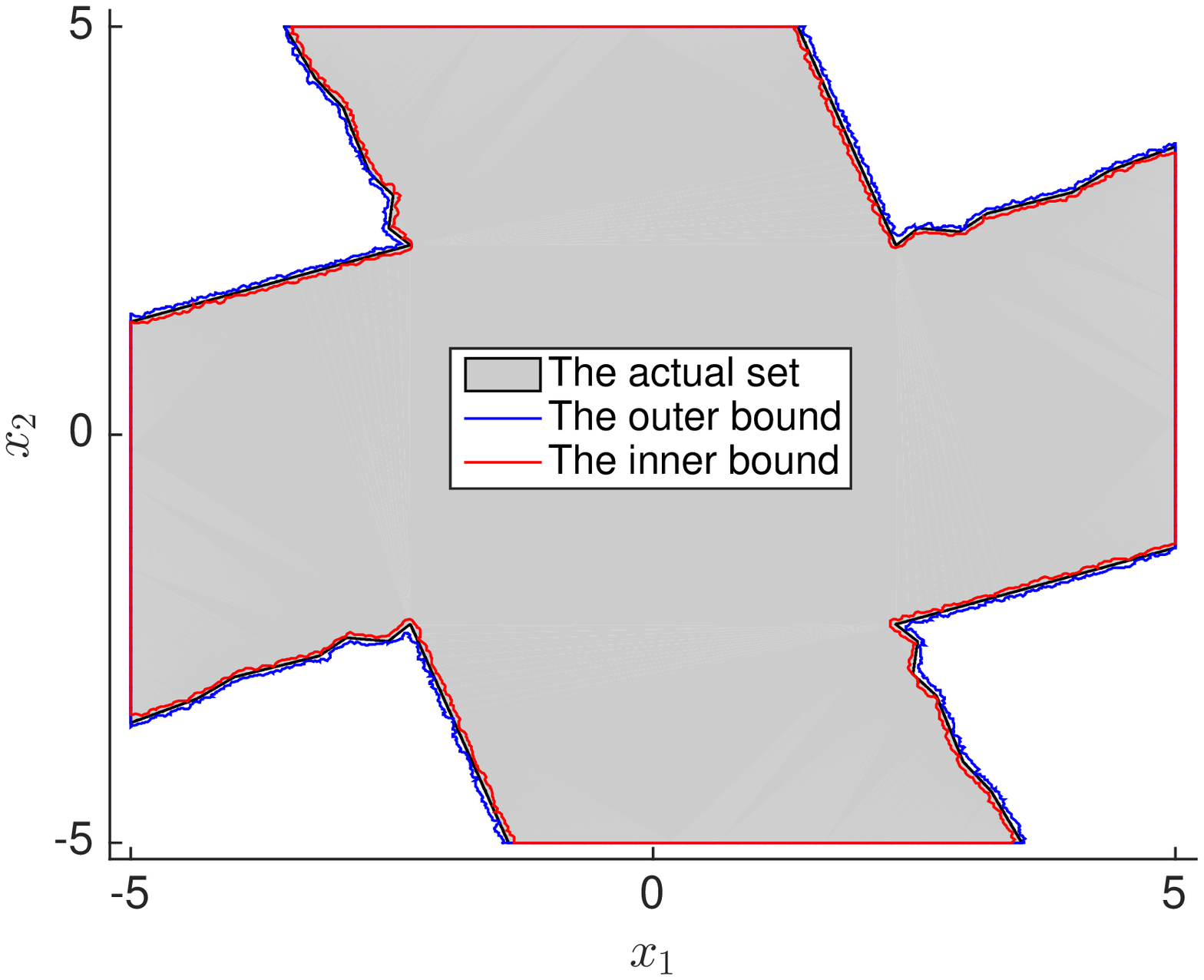}}\\
  \end{tabular}
\caption{Set identification for two different values of $\bar{\delta}$ in Example \ref{exam:pwa}: (a) $\bar{\delta}=0.1$ and (b) $\bar{\delta}=0.05$. }
\label{fig:pwa}
\end{figure}

\begin{example}\label{exam:pwa}
Finally, we will present an example where the dynamics are not continuous. We consider the following piecewise affine (PWA) system
\begin{align*}
x(t+1) = A_{\sigma(x(t))}x(t), \quad \sigma(x) =  \begin{cases}
      1 & \textrm{if } |x_1|>|x_2|,\\
      2 & \textrm{if } |x_1|\le|x_2|,\\
   \end{cases}
\end{align*}
where $A_1 = e^{A_1^C}$ and $A_2 = e^{A_2^C}$ with
\begin{align*}
A_1^C = \begin{bmatrix}
-0.1 & 5\\
-1 & -0.1
\end{bmatrix}, ~ A_2^C = \begin{bmatrix}
-0.1 & 1\\
-5 & -0.1
\end{bmatrix}.
\end{align*}
\end{example}
This example is a discretized system of Example 1 in \cite{ART:JA98} and it is asymptotically stable. However, the dynamics are not continuous. The state constraints are given by $X=\{x\in \mathbb{R}^2: |x_1|\le 5, |x_2|\le 5\}$. Under the same setting in Example \ref{exam:nl}, the solution of $\mathcal{P}^*(\omega_N)$ is $t^*(\omega_N)=4$. For two different values $\bar{\delta}$, the inner and outer approximations are presented in Figure \ref{fig:pwa}.

\section{Conclusions}\label{sec:con}
We have presented a data-driven framework to compute the maximal invariant set of discrete-time black-box nonlinear systems by using a finite number of trajectories. Our approach relies on the mathematical theory of set invariance in control, the scenario approach, and the recently introduced notion of almost-invariant sets. Our assumptions are very mild and standard. Despite this, we show that one can compute almost-invariant sets using the proposed approach and that  probabilistic invariance guarantees of the computed set can be established. Our approach only requires that one can simulate the system with a priori fixed initial conditions. For an explicit expression of the almost-invariant set obtained from our approach, we have also developed a data-driven set identification procedure that gives inner and outer approximations within a prescribed tolerance and confidence level. Finally, we have demonstrated the applicability of the proposed data-driven framework on several complex systems.

\appendix
\section*{An alternative bound for Theorem 1}
It is also possible to derive a bound using Hoeffding's inequality in Theorem 1. For interested readers, we also present this bound. However, it is looser than the one in Theorem 1.

From the definition of $\theta_k(\omega_N)$ in (\ref{eqn:ttheta}), we can see that
\begin{align}
\theta_k(\omega_N)-\theta_{k+1}(\omega_N) = \frac{\sum\limits_{x\in \omega_N}\pmb{1}_{O_{k}\setminus O_{k+1}}(x)}{N}
\end{align}
For notational convenience, let $\Delta \theta_k(\omega_N) = \theta_k(\omega_N)-\theta_{k+1}(\omega_N)$ for all $k\in \mathbb{Z}^+$. Hence, from Hoeffding's inequality, 
\begin{align}
\mathbb{P}^N(\omega_N\in X^N:|\Delta \theta_k(\omega_N)-S(k) | \ge \epsilon)& \le 2e^{-2N\epsilon^2}. \label{eqn:Delatathetak}
\end{align}
for all $k\in \mathbb{Z}^+$. Again, any $\epsilon\in (0,1]$, we consider the set $\mathcal{I}_{\epsilon} = \{k\in \mathbb{Z}^+: S(k) \ge \epsilon\}$. The set $\{\omega_N\in X^N:S(\bar{t}(\omega_N))\ge \epsilon\}$ can be written as $\cup_{k\in \mathcal{I}_{\epsilon} }\{\omega_N\in X^N: \bar{t}(\omega_N) = k \}$. For each $k\in \mathcal{I}_{\epsilon}$, $\{\omega_N\in X^N: \bar{t}(\omega_N) = k \}$ can be rewritten as $\{\omega_N\in X^N:|\Delta \theta_k(\omega_N)-S(k) | \ge \epsilon, \bar{t}(\omega_N) = k \}$, whose measure is bounded by $2e^{-2N\epsilon^2}$  from (\ref{eqn:Delatathetak}). Following the same argument in the proof of Theorem 1, the measure of the set $\{\omega_N\in X^N:S(\bar{t}(\omega_N))\ge \epsilon\}$ is bounded by $\frac{2}{\epsilon}e^{-2N\epsilon^2}$.  $\Box$

\bibliographystyle{unsrt}
\bibliography{Reference}

\end{document}